\documentclass[journal]{IEEEtran}

\usepackage{cite}
\usepackage{amsmath,amssymb,amsfonts}
\usepackage{algorithmic}
\usepackage{graphicx}
\usepackage{textcomp}

\usepackage{cite}
\usepackage{amsmath,amssymb,amsfonts}
\usepackage{algorithmic}
\usepackage{graphicx}
\usepackage{epstopdf}
\usepackage{pgfplots}
\usepackage{subfigure}
 \graphicspath{{\figures}}
\usepackage{textcomp}
\usepackage{xcolor}
\usepackage{verbatim}
\usepackage{makecell}
\usepackage{booktabs}
\usepackage{CJK}
\usepackage{url}

\ifCLASSINFOpdf
\else
\fi

\pgfplotsset{compat=1.14}

\begin{document}

\title{Mobility-Aware Seamless Handover with MPTCP in Software-Defined HetNets}
%
%
%
\author{Haonan~Tong,~Tao Wang,~Yujiao Zhu,~Xuanlin Liu,~Sihua Wang,
        \\ ~and Changchuan Yin,~\IEEEmembership{Senior Member,~IEEE}
\thanks{H. Tong, T. Wang, Y. Zhu, X. Liu, S. Wang, and C. Yin are with the Beijing Key Laboratory of Network System Architecture and Convergence, and also with the Beijing Advanced Information Network Laboratory, Beijing University of Posts and Telecommunications, Beijing, 100876 China (e-mail: hntong@bupt.edu.cn, taowang@bupt.edu.cn, yjzhu@bupt.edu.cn,  xuanlin.liu@bupt.edu.cn,  sihuawang@bupt.edu.cn,  ccyin@bupt.edu.cn).}
}   


%
%

\markboth{}%
{H.Tong \MakeLowercase{\textit{et al.}}: Mobility-Aware Seamless Handover with MPTCP in Software Defined HetNets}

%



\maketitle

\begin{abstract}
In this paper, the problem of vertical handover in software-defined network~(SDN) based heterogeneous networks~(HetNets) is studied. In the studied model, HetNets are required to offer diverse services for mobile users. Using an SDN controller, HetNets have the capability of managing users' access and mobility issues but still have the problems of ping-pong effect and service interruption during vertical handover. 
To solve these problems, a mobility-aware seamless handover method based on multipath transmission control protocol~(MPTCP) is proposed. \color{black}
The proposed handover method is executed in the controller of the software-defined HetNets~(SDHetNets) and consists of three steps: location prediction, network selection, and handover execution.
In particular, the method first 
predicts the user's location in the next moment with an echo state network~(ESN).
Given the predicted location, the SDHetNet controller can determine the candidate network set for the handover to pre-allocate network wireless resources.
Second, the target network is selected through fuzzy analytic hierarchical process~(FAHP) algorithm, jointly considering user preferences, service requirements, network attributes, and user mobility patterns.
Then, seamless handover is realized through the proposed MPTCP-based handover mechanism.
\color{black}
Simulations using real-world user trajectory data from Korea Advanced Institute of Science \& Technology show that the proposed method can reduce the handover times by 10.85\% to 29.12\% compared with traditional methods. The proposed method also maintains at least one MPTCP subflow connected during the handover process and achieves \color{black} a seamless handover. 
\color{black}

\end{abstract}


%
\IEEEpeerreviewmaketitle

\section{Introduction}
\IEEEPARstart{T}{he}  next-generation mobile networks will be more user-centric and will be able to carry diversified services. In addition, multiservice multimode terminals~(MMTs) running multiple services simultaneously will become ubiquitous~\cite{handover_servey}. To satisfy various quality-of-service~(QoS) requirements, heterogeneous networks were introduced for multiple user accesses~\cite{het}. Because network access points with a higher carrier frequency have a smaller coverage area, vertical handovers, which change the type of connected network, have become pervasive.
However, vertical handovers across HetNets still face many challenges that range from user mobility perception to network selection and service continuity guarantee.  
Since the emerging software-defined network~(SDN) is standardized to be equipped with access and mobility management function~(AMF) in the control plane~\cite{3GPP}, vertical handovers in SDHetNets can be further improved by saving network resources and guaranteeing service continuity.

\vspace{-0.3cm}
\subsection{Related Works}
\color{black}In recent works, the location prediction of mobile users provided key information for network discovery and avoiding ping-pong effect. \color{black}
In~\cite{HMM}, the authors proposed a hidden Markov model-based algorithm to predict the MMT's future location, thus reducing ping-pong effect during handover. 
The work in~\cite{Claster} introduced a cluster-aided mobility algorithm to predict user location  by exploiting similarities among users’ mobility.
The work in~\cite{Cmeans} introduced the sequence pattern mining method to give users' next area predictions with a crowd's statistical trajectory features, where the user mobility patterns in different time periods are extracted for location prediction.
The authors in~\cite{Cacheinsky} proposed a conceptor-based echo state network~(ESN) framework that predicted the user mobility pattern to proactively allocate cache content for users in the human-in-the-loop scenario. However, all of these works~\cite{HMM,Claster,Cmeans,Cacheinsky} predicted user locations with long-term mobility patterns or crowd movement trends, which are not capable of recognizing the transitory moving patterns for network handover. In this case, a predictive model with a transitory location prediction is required to recognize the user's ping-pong moving pattern.

\color{black}Current studies on network selection for handover \cite{game_theory1,game_theory2,newest_journal,select_ML1,select_ML2,tutorial,select_Qlearning,select_fuzzy_logic_sys,select_fuzzy_logic2,select_attrib_user_prefer,MCGDM,Utility-GDM,select} mainly focused on guaranteeing multiservice QoS requirements and satisfying user preferences for diverse services. \color{black}
The works in \cite{game_theory1,game_theory2} constructed the network selection problem as a game theory model, where mobile users selected the target network based on the utility function considering several factors, such as user's utility and channel allocation. \color{black}However, the game theory methods assumed that users compete for the bandwidth of diverse networks, which cannot guarantee the algorithm's convergence~\cite{newest_journal}. \color{black}
In~\cite{select_ML1,select_ML2,tutorial,select_Qlearning}, machine learning algorithms, especially $Q$-learning~\cite{select_Qlearning}, were introduced to make network selection decisions by learning from the environment. However, these algorithms are not robust enough to adapt to dynamic network fluctuations, which are not suitable for high-speed moving scenarios.  \color{black}
The authors in~\cite{select_fuzzy_logic_sys,select_fuzzy_logic2} used a fuzzy logic-based inference engine to select a network during handover, which could reduce ping-pong effect with if-then-else rules. However, the values of network attributes were not quantified. 
The works in~\cite{select_attrib_user_prefer,MCGDM,Utility-GDM,select} used group decision making methods to determine the weights of different attributes for decisions. The work in~\cite{select} proposed a comprehensive network selection algorithm based on the fuzzy analytic hierarchy process~(FAHP), which considered both user preferences and service QoS requirements.
Note that, in~\cite{select_attrib_user_prefer,MCGDM,Utility-GDM,select}, the network selection computation is deployed on the MMT, which
 has a huge handover delay when network selection algorithms are complex. Additionally, selecting a network on the MMT neglects the users' mobility perception in the handover process. In this case, SDN controller with great computational power can be considered to undertake network selection computation according to~\cite{2interface,speedup}. 

\color{black}
The works in existing literature~\cite{anchor_based_handover,wen_xiang_ming,MPTCP,acmmptcp,Tsinghua,tong_hao_nan,phone, mptcp2014 } studied handover mechanism for service continuity.
\color{black}
The work in~\cite{anchor_based_handover} proposed an anchor-based multi-connectivity mechanism to reduce the handover cost in 5G network. 
The work in~\cite{wen_xiang_ming} realized unchanged IP during handover with a virtual interface in SDN.
However, these works
used only one interface during the vertical handover, which requires link reconnection or data migration, thus inevitably bringing service delays, or even breaks. 
Alternatively, in~\cite{MPTCP} the authors discussed MPTCP as a potential solution for mobility-related service continuity issues, and several works~\cite{MPTCP,acmmptcp,tong_hao_nan, Tsinghua}~investigated MPTCP-based seamless handover. 
Early works on MPTCP~\cite{phone, mptcp2014} focused on improving MPTCP  throughputs in static scenarios without considering handovers. 
In \cite{acmmptcp}, the authors analyzed MPTCP-based vertical handover in a fixed handover mode.
In recent years, using SDN controller's capability of managing subflows, MPTCP has been tested for seamless handover.
The works in~\cite{Tsinghua,tong_hao_nan} verified the feasibility of MPTCP-based handover methods in mobile scenarios, where MPTCP had to establish a new dataflow connection with a new network. \color{black}However, these works did not clearly illustrate the signaling process or path management strategies for MPTCP during handover, especially the process of how MPTCP establishes a new connection and migrates data transmission to the target network.
\color{black}
\color{black}

\vspace{-0.3cm}
\subsection{Contributions}

The main contribution of this paper is that we propose a novel \color{black} mobility-aware \color{black} seamless handover method with MPTCP in SDHetNets to avoid ping-pong effect and guarantee service continuity during a vertical handover. Specifically, our key contributions include:
\begin{itemize}
    \item We consider an SDHetNet, in which a vertical handover meets problems with  ping-pong effect and service continuity. 
    The controller of SDHetNet must handle the user's transitory mobility information, select the target network, and instruct handover execution.
    \item To solve the problems of ping-pong effect and service continuity, we propose a modified ESN model to predict the user's location in the next moment. Using historical locations, the model can extract the user's transitory moving pattern and provide accurate predictions of the user locations, thus determining the candidate network set in a proactive manner.
    \item Given the candidate network set, we develop an FAHP-based network selection algorithm to determine the target network with the highest Quality-of-Experience~(QoE) metric, which synthetically considers user preferences, QoS requirements, and the network's real-time attributes. Then, we propose a seamless handover mechanism that uses MPTCP to transmit data through multiple networks, thus guaranteeing service continuity during handover.
    \item \color{black}
    We perform a fundamental analysis on the performance of the proposed method. 
    The simulation results show that the ESN model can handle the multivariate time series prediction problem by offline fast training with small data samples. Meanwhile, the proposed FAHP based network selection algorithm can select the appropriate network and efficaciously avoid ping-pong effect. Then, the throughput results show that the proposed MPTCP-based handover mechanism can maintain the subflow connected during handover, \color{black}thus achieving a seamless handover.
    \color{black}
\end{itemize}

Simulations use real-world data from the Korea Advanced Institute of Science \& Technology~(KAIST) as the user trajectory.
The simulation results show that, compared with the benchmark network selection algorithms, the handover times of the proposed method can be reduced from 10.85\% to 29.12\%, and that the service continuity is also guaranteed during handover.
\emph{To the best of our knowledge, this is the first work to analyze the seamless handover in SDHetNets, given the user location prediction and network QoE metrics calculated by  SDN controller.}

The remainder of this paper is organized as follows. 
The system model and problem formulation are proposed in Section II. 
In Section III, we give a detailed description of the proposed seamless handover method.
The simulation results are presented and analyzed in Section IV.
The conclusions are given in Section V.



\begin{figure}[t]
    \centering
    \includegraphics[width=0.42\textwidth,height = 0.35\textwidth
    ]{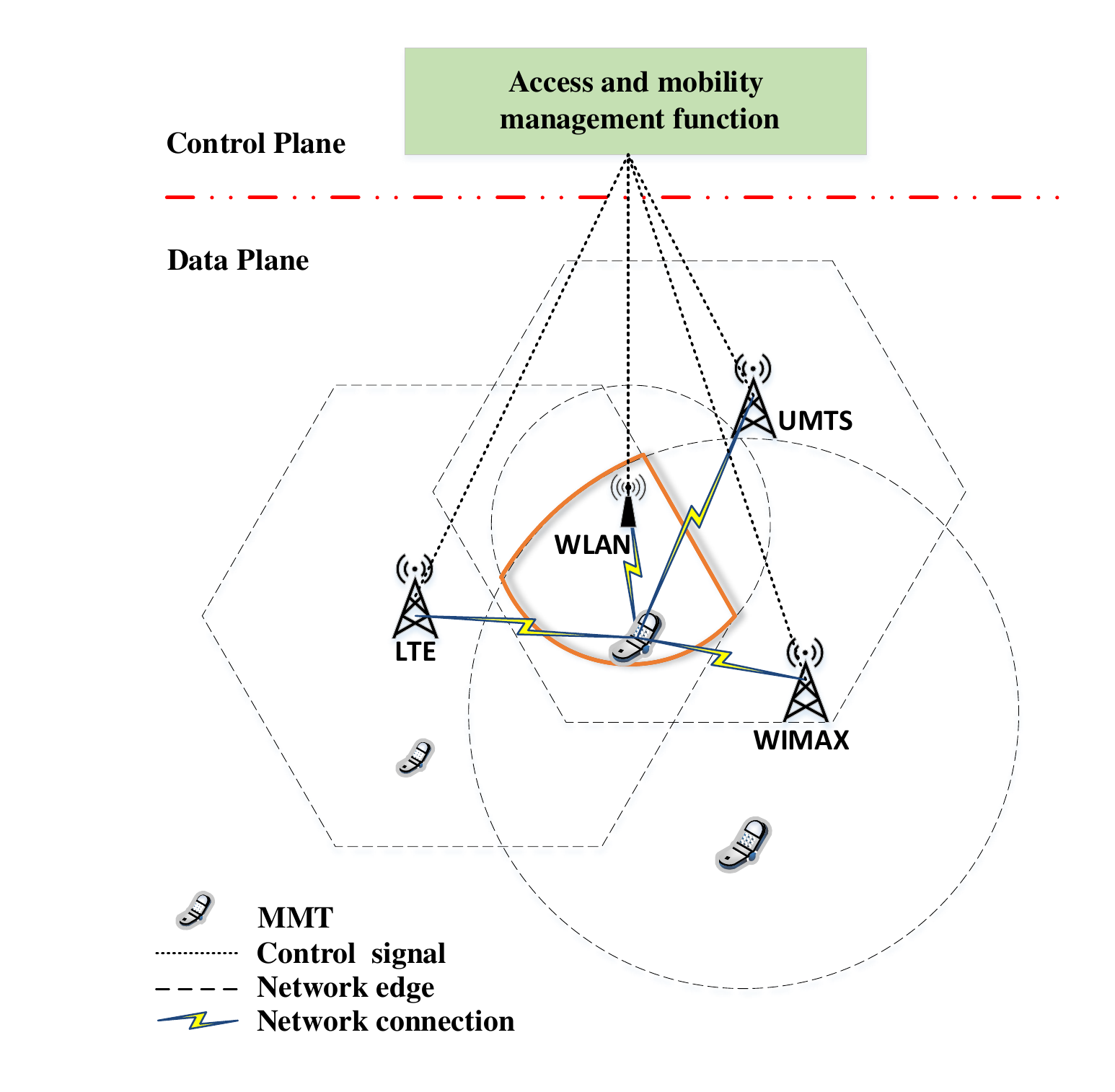}
    \setlength{\abovecaptionskip}{0.cm}
    \caption{The architecture of SDHetNet.}\label{fig:HWN}
    \vspace{-0.6cm}
\end{figure}

\section{System Model and Problem Formulation}

In this paper, SDHetNet is functionally separated into a control plane and a data plane according to the architecture of SDN~\cite{3GPP}. 
In the SDN control plane, SDN controller is equipped with AMF module that has the capability of managing users' access and mobility. 
\color{black}
In  the SDN data plane, multiple types of network access points and SDN switches exist, and the data packets are routed and forwarded by switches under the instructions from the controller using OpenFlow protocol~\cite{3GPP}. \color{black}
In this system, we assume that the seamless network handover is completed in cooperation with the SDN controller. The SDN controller is configured to collect, store, and process the data from both users and networks. In this case, the SDN controller can predict the user's location for network discovery, run the network selection algorithm, and instruct the handover process with control signaling.

We consider the SDHetNet scenario consists of $M$ types of networks, including Universal Mobile Telecommunications System (UMTS), World Interoperability for Microwave Access (WiMAX), Long Term Evolution (LTE), and Wireless Local Area Network (WLAN), as shown in Fig.~\ref{fig:HWN}. 
In the SDHetNet, \color{black}handovers are controlled by one unified SDN controller, \color{black} $U$ mobile users with MMTs move casually, and the MMTs can connect to any type of network when the MMTs are located within the network's range. 
\color{black}Each MMT can periodically upload the location coordinates $\boldsymbol m_{t,j}$ of user $j$ to the SDN controller in an interval of $H$ seconds. 
\color{black}
Collecting the user's ${N_{in}}$ historical locations, the SDN controller predicts the future location $\boldsymbol{s}_{t,j}$ of user $j$ in each period to decide the candidate network set $\mathcal{R} = \{r_1,\dots,r_M \}$ in a proactive manner.
Then, the target network $r_t$ for handover is calculated by SDN controller, such that the network selection delay can be decreased. 
Concretely, $r_t$ is selected from $\mathcal{R}$, maximizing the QoE metric of the network, which combines $K$ types of user preferences, $L$ types of services' QoS requirements, and $N$ network attributes.
Given $r_t$, a seamless handover execution can be achieved with SDN controller controlling MPTCP connections.
During the handover execution, SDN controller instructs the MMT to establish a new MPTCP session that connects to both the originally connected network and the target network. In this case, the data packets are alternatively transmitted through multiple network connections, thus avoiding data reforward and guaranteeing service continuity. 
\color{black}Table~\ref{tab:notations} provides a summary of the notations used in this paper. \color{black}Then, we present the mobility model, transmission model, QoE model, and problem formulation.

\begin{table}[t]
    \centering
    \setlength{\abovecaptionskip}{-0.6cm}
    \caption{List of Notations}
    \begin{center}
        \begin{tabular}{|c | c |}
        \cline{1-2}
            \hline
            \textbf{Notation}&\textbf{Description} \\
            \hline
            $M$ & Number of network types         \\
            \hline
            $U$ & Number of users \\
            \hline
            $K$ & Number of user preference types   \\
            \hline
            $L$ & Number of service types     \\
            \hline
            $N$ & Number of attributes considered for network selection    \\
            \hline
            ${H}$ & Duration of prediction period  \\ 
            \hline
            $N_{in}$    &  Number of historical locations used for location prediction  \\
            \hline
            $N_s$    &  Number of predicted locations   \\
            \hline
            $\boldsymbol m_{t,j}$ & Location vector with $N_{in}$ historical locations of user $j$  \\ 
            \hline
            $\boldsymbol s_{t,j}$ & Predicted location vector  with $N_s$ locations of user $j$\\           
            \hline
            
            $\boldsymbol{W}^{\mathrm{S}}_{l}$  & Weight vector of network attributes service $s_l$       \\
            \hline
            $\boldsymbol P_k$ &  User service preference vector    \\
            \hline
            $\boldsymbol{A}_{t}$ & Collected real-time network attribute values    \\
            \hline
            $S_i$ & QoE value of network $r_i$    \\
            \hline

        \end{tabular}
        \label{tab:notations}
    \end{center}
    \vspace{-0.6cm}
\end{table}

\vspace{-0.2cm}
\subsection{Mobility Model}
    In the proposed SDHetNet, we consider that the users move continuously. During the day, the user with MMT may stay in a small area or move a long distance.
    Each MMT is authenticated to periodically upload the location coordinates $\boldsymbol m_{t,j}$ to SDN controller in intervals of $H$ seconds. 
    In each period, the mobility-aware SDN controller uses ${N_{in}}$ collected historical locations to extract the users' mobility patterns.\color{black} Thus, the SDN controller can proactively predict the MMT's locations $\boldsymbol s_{t,j}$ in the next period.
    Then, in this model, the predicted locations are used to determine the set of candidate access networks.

    Given $M$ types of networks in SDHetNet, we assume that the MMT can connect to any type of network as long as the MMT is in the coverage area of the corresponding access points.
    For each type of network, we select the nearest access point to the predicted location as one element in the set of candidate access networks for handover.
    The candidate network set is defined as $\mathcal{R} = \{r_1,\dots,r_M \}$, where $r_m$ denotes the $m$-th type of network. 
\subsection{Transmission Model}
    Here, we introduce the models for transmission links between the MMT and the associated network access points. 
    For tractability, we uniformly denote the base stations and WLAN APs  as network access points~\cite{FMC}.
    The received signal strength~(RSS) is defined as a vital attribute for network selection.
   \color{black}For networks with macro base stations, such as UMTS, WiMAX, and LTE, the MMT's RSS is obtained through cost 231-Hata model~\cite{path_loss}. Taking the transmission path loss into account, the RSS is represented as~(in dBm):
   \begin{equation}
       \label{equ:hata231}
    \color{black} {P}_{rB} = {P}_{tB} - 127.5 -35.2\lg d_B - {x}_{\sigma B},\\
   \end{equation}
   where ${P}_{tB}$ is the transmitted power of the macro base station in dBm, $d_B$ indicates the distance between the MMT and the macro base station in km, and $x_{\sigma B}$ is a zero mean Gaussian distributed random variable with standard deviation $\sigma B$, indicating shadow fading caused by obstacles. For WLAN, we take the free space loss into account, and the RSS from WLAN AP is represented as\cite{path_loss}~(in dBm):
   \begin{equation}
       \label{equ:wlanloss}
    \color{black}  {P}_{rW} = {P}_{tW} -35.2 - 20 \lg f_W - 20\lg d_W - {x}_{\sigma W},
   \end{equation}
   where ${P}_{tW}$ is the transmitted power of WLAN AP in dBm, $f_{W}$ denotes the carrier frequency in MHz, $d_W$ denotes the distance between the MMT and AP in km, and $x_{\sigma W}$ is a zero mean Gaussian distributed random variable with a standard deviation $\sigma W$, indicating shadow fading caused by obstacles.\color{black}

\subsection{Quality-of-Experience Model}
In the proposed SDHetNet, the SDN controller can collect user preferences for services, diverse QoS requirements and real-time network attributes. In this subsection, we define the QoE metric as a map of the QoS metric and jointly consider 
supplementary parameters.


Given the candidate network set $\mathcal{R}$,  assume that the MMTs can run a set of $Y$ types of services $\mathcal{S}=\left\{s_1,\ldots,s_{L} \right\}$. Different services have distinguished requirements for network attributes. 
In the network selection stage, a  set of network attributes $\mathcal{C}=\left\{c_1,\ldots,c_N\right\}$ is considered as the criteria, where $N$ denotes the number of considered network attributes, such as RSS, bandwidth, and delay. To discriminate various QoS requirements for network attributes, the weight vector~$\boldsymbol{W}^{\mathrm S}_{l} = [{w}^{\mathrm S}_{l,1}, {w}^{\mathrm S}_{l,2},\dots {w}^{\mathrm S}_{l,N}]$ of the network attributes is introduced, where $ {w}^{\mathrm S}_{l,n} $ denotes the weight of attribute $c_{n}$ for service $s_{l}$. Considering that mobile users have different preferences among $L$ types of services, the service priority vector $\boldsymbol{P}_{k} = [p_{k,1},p_{k,2},\dots,p_{k,L}]$ is defined to represent the user's preference, where $p_{k,y}$ indicates the priority of service $s_l$ for the $k$-th type of preference.

Collected periodically by SDN controller, the real-time attributes $\boldsymbol{A}_t$ of $M$ types of networks are represented in a matrix form:
\begin{equation}   
    \label{equ:real-time attrib}
    \boldsymbol{A}_t = {({a_{ij}})_{M \times N}}  ,
\end{equation}
where $a_{ij}$ indicates the collected raw value of network $r_i$ for attribute $c_j$. 
Then the network attributes are normalized with utility functions.   
The normalized attribute utility value is denoted as  $\boldsymbol{U}^i_l = \left[ {u_{l,1}^{i},u_{l,2}^{i}, \ldots ,u_{l,N}^{i}} \right]$, where~$\boldsymbol{U}^i_l$ represents $r_i$'s utility vector for service $s_l$.
Combining user preferences, QoS requirements and real-time network attributes, the QoE metric of network $i$ is defined as~\cite{select}:
 \begin{equation}    
 \label{equ:QoE}
 {S_i} = \sum_{l = 1}^L {{p_{k,l}}}  \cdot \boldsymbol{W}^{\mathrm{S}}_{l} \cdot {\boldsymbol{U}_l^i}^T.
 \end{equation}

\subsection{Problem Formulation}
The purpose of network selection is to determine the network with the highest QoE value. The problem can be formulated as a rank problem. Given the user preferences $\boldsymbol{P}_k$ and network attributes $\boldsymbol{A}_t$ collected by SDN controller, the target network $r_t$ is defined as:
     \begin{equation}
     \label{equ:argmax}
           {t} = \mathop {{\mathop{\mathrm{arg~max} }\nolimits} }\limits_{t \in \left\{1 \cdots M\right\}} {\kern 1pt} {S_t} ,
     \end{equation} 
where $t$ is the index of the target network.
Here, we note that the user's location information is used to determine the candidate network set $\mathcal{R}$.
The weights of the attributes are obtained from the relative importance between every two attributes given in a pairwise way. We apply FAHP to quantify the weight vectors $\boldsymbol{W}^{\mathrm{S}}_{l}$. The details will be expounded in section III. Given the target network for a potential handover, handover execution should be triggered on certain conditions to avoid ping-pong effect.

Given the target network, MPTCP is introduced to realize seamless handover execution. 
In SDHetNets, MPTCP has the capability of simultaneously using multiple paths to transmit data between peers under the instruction of SDN controller.
During the handover, when the quality of the originally connected network's link decays while another network's link is available, the handover conditions are triggered.
   SDN controller will send the MMT handover control signaling  to initiate a new MPTCP session, which establishes connections with multiple available networks
   and diverts the data transmission from the original MPTCP session to the new one. The original MPTCP session will be terminated after the new MPTCP session is established. 
   In the handover execution process, the MPTCP connections maintain the throughputs when changing networks, and the handover delay is diminished.
   In this way, the service continuity during handover is guaranteed, thus realizing seamless handover in SDHetNets.

\section{Seamless handover method }
    In this section, we propose the \color{black} mobility-aware \color{black} seamless handover method in the considered SDHetNet. 
    To complete the handover process, we first predict the user's location in a short future time and determine a set of candidate network access points as the discovered networks. 
Then, in the network selection stage, and given the candidate network set, we apply FAHP to quantify the attribute weights.
The target network is selected with the highest QoE metric from the candidate network set. Meanwhile, with the collected user's historical locations, the mobile user's ping-pong moving pattern can be recognized to avoid reiterative handovers. 
     After that, the MMT's handover execution to the target network is instructed by SDN controller, and the MMT establishes MPTCP subflows to the originally connected network and the target networks, thus maintaining a subflow connection during handover.  
     In the following subsections, we introduce the location prediction, the FAHP-based network selection algorithm, and MPTCP-based seamless handover execution in detail.

\begin{figure}[htbp!]
    \vspace{-1.0cm}
    \centering
    \includegraphics[width=0.45\textwidth, height=0.3\textwidth
    ]{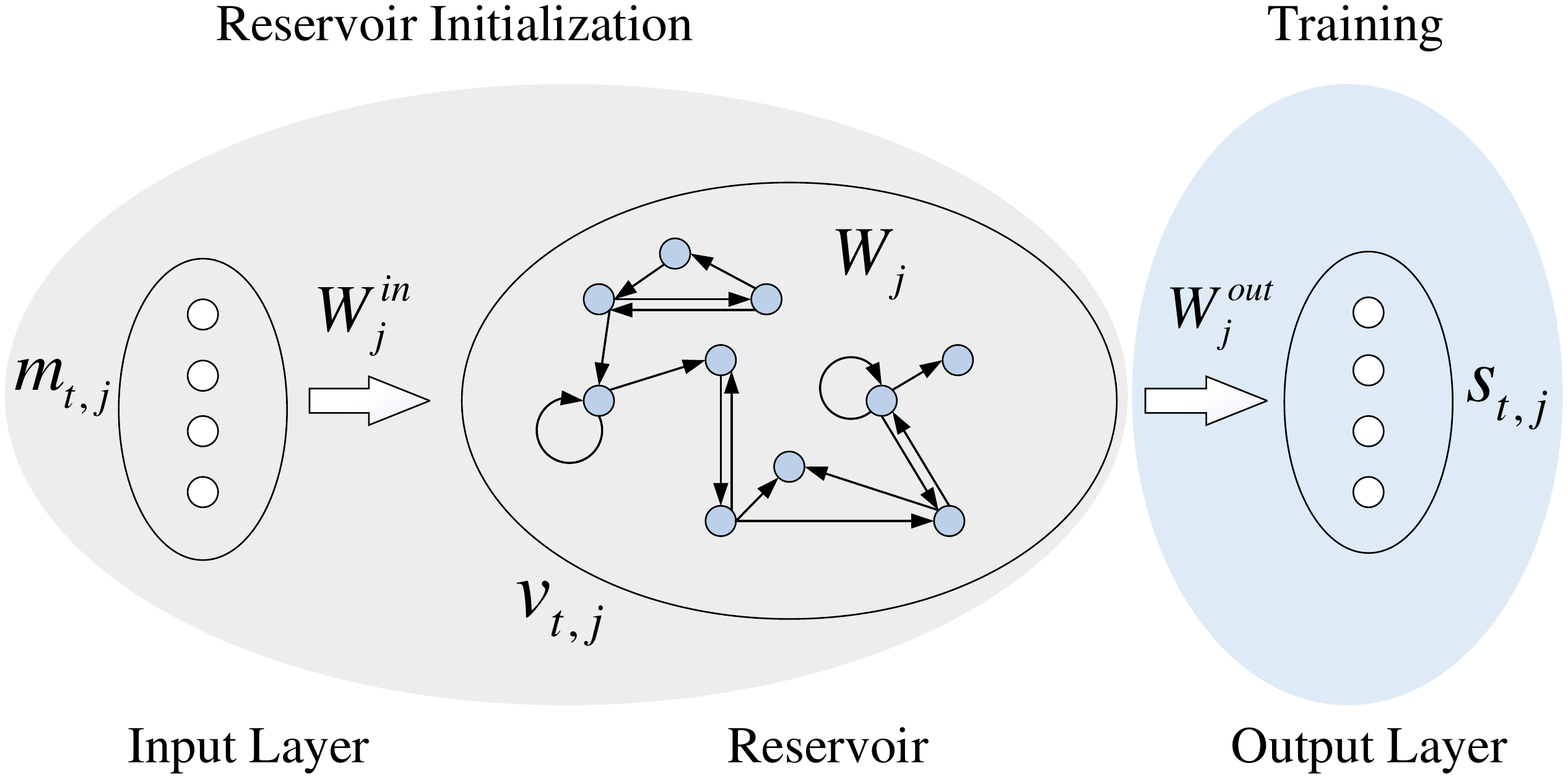}
    \setlength{\abovecaptionskip}{-0.6cm}
    \caption{Structure of echo state network and training approach.}\label{fig:esn}
    \vspace{-0.6cm}
\end{figure}

\subsection{Location Prediction}
In this subsection, we propose a location prediction algorithm  to decide the candidate networks in the network discovery stage.
\color{black}
As user locations are indeterminate, it is necessary to predict the locations for handover learning from users' historical mobility patterns.
To extract mobile users' instantaneous mobility patterns, we introduced a recurrent neural network~(RNN) model, ESN, which learns from users' historical locations and then infers future locations. \color{black} 
The ESN architecture shown in Fig.~\ref{fig:esn} is based on a neuron reservoir, in which the recurrent neurons are randomly connected. The reservoir is driven by a transitory input, and the
activation state induced by previous input signals can echo in the reservoir, thus, the state of the reservoir is a rich representation of the long and short-term memory of the inputs. In this case, a simple linear combination of the reservoir units is a good predictor of future inputs.
Unlike other prediction algorithms in~\cite{HMM,Claster,Cmeans}, 
the ESN reservoir loads the transfer relationship among locations as well as the user's instantaneous movement state, which is more suitable for transitory prediction than Markov models.  
Moreover, ESN is trained in a quick offline manner with small data samples, which makes it more flexible for deployment in distributed network controllers than other RNN models~\cite{LSTM}.

In our model, ESN predicts the location coordinates in the next moment based on the historical locations of the user. Therefore, during the network training process, we choose the offline method to train the output matrix. The location prediction approach is composed of three parts: input, output, and ESN model. The specifics of the location prediction approach are defined as follows:

$\bullet\;$\textbf{\emph{Input}}: 
The two-dimensional location coordinates of user $j$ at time $t$ are denoted as $m_{t,j}=(x_{t,j},y_{t,j})$, where $x_{t,j}$ and $y_{t,j}$ represent the two-dimensional location coordinates. 
The input variables can be expressed as a $2{N_{in}} \times 1$ vector, that is, $\boldsymbol m_{t,j} = [m_{t,j},\dots,m_{t-N_{in}+1,j}]^T$. Here, ${N_{in}}$ represents the
number of historical location data points in the input variables. 

$\bullet\;$\textbf{\emph{Output}}: 
The output of ESN model is defined as $\boldsymbol s_{t,j}=[s_{t,j1},\dots,s_{t,jN_s}]^T$, where $\boldsymbol s_{t,j}$ denotes the predicted $N_s$ locations of user $j$ to visit in the next prediction period, and the prediction period is set as $H$ seconds.


$\bullet\;$\textbf{\emph{ESN Model}}: An ESN model can predict the user's future location coordinates based on historical data; that is, the output variable $\boldsymbol s_{t,j}$ can be obtained from the input variable $\boldsymbol m_{t,j}$. The ESN model contains three matrices: the input weight matrix $\boldsymbol{W}_j^{in}\in\mathbb{R}^{W \times2N_{in}}$, the recurrent matrix $\boldsymbol{W}_j\in\mathbb{R}^{W\times W}$, and the output matrix $\boldsymbol{W}_j^{out}\in\mathbb{R}^{2N_s\times W}$, where $W$ represents the number of neurons in the reservoir.
To reduce the uncertainty and improve the utilization of neurons, in our model the recurrent matrix $\boldsymbol{W}_j$ is defined as a full rank sparse matrix according to~\cite{esn_cycle_jumps}: 

\begin{equation}   
\label{equ0}
\boldsymbol{W}_j=\left[
\begin{matrix}
0&~&~&~&w&0&w\\
w&0&~&~&~&0&0\\
0&w&0&~&~&~&w\\
\ddots&\ddots&\ddots&\ddots&~&~&~\\
~&w&0&w&0&~\\
~&~&0&0&w&0\\
~&~&~&w&0&w&0\\
\end{matrix}
\right],
\end{equation}
where $w$ in the matrix obeys Gaussian distribution. 
In this way, the neurons in the reservoir are connected in a cycle and have jumping connections in every two neurons~\cite{esn_cycle_jumps}.
Based on the above conditions, the reservoir state variable $\boldsymbol v_{t,j}$ of user $j$ at time $t$ can be updated according to a nonlinear method. The calculation formula is:
\begin{equation}   
\label{equ200}
\boldsymbol{v}_{t,j}= \mathrm{tanh}(\boldsymbol{W}_j\boldsymbol{v}_{t-1,j}+\boldsymbol{W}_{j}^{in}\boldsymbol m_{t,j} ) .
\end{equation}

Then according to the output matrix $\boldsymbol{W}_j^{out}$ and the reservoir state variables $\boldsymbol{v}_{t,j}$, the output variable $\boldsymbol{s}_{t,j}$ can be calculated as follows:

\vspace{-0.3cm}
\begin{equation}  
\label{equ102}
\boldsymbol{s}_{t,j}=\boldsymbol{W}_{j}^{out}\boldsymbol{v}_{t,j} .
\end{equation} 

The input matrix $\boldsymbol{W}_j^{in}$ and the recurrent matrix $\boldsymbol{W}_j$ of the ESN are determined during the initialization process and will not change, 
and the output matrix $\boldsymbol{W}_j^{out}$ is trained in an offline manner using ridge regression~\cite{ridge_regression}:

\begin{equation}   
\label{equ103}
\boldsymbol{W}_{j}^{out}=\boldsymbol{s}_j\boldsymbol{v}_j^{T}(\boldsymbol{v}_j^{T}\boldsymbol{v}_j+\lambda^2\boldsymbol{I})^{-1} ,
\end{equation} 
where $\boldsymbol{v}_j=[\boldsymbol{v}_{1,j},\dots,\boldsymbol{v}_{N_{tr},j}]\in\mathbb{R}^{W \times N_{tr}}$ represents the reservoir states of user $j$, and $\boldsymbol{s}_{j} \in \mathbb{R}^{2N_s \times N_{tr}}$ is the combination of $N_{tr}$ outputs within a period, and $\boldsymbol{I}$ is the identity matrix. 
After that, the output $\boldsymbol{s}_{t,j}$ can be obtained with the trained model to infer the mobile users' future locations in an online manner. 

Given the predicted user location in the next moment, SDN controller selects one from each type of network access point to form the candidate network set.
The candidate access point is selected as the closest access point to the predicted location thus simplifying the searching space of the network selection algorithm.

\vspace{-0.2cm}
\subsection{FAHP-Based Network Selection Algorithm}

In this subsection, we use the candidate network set to select the target network in the network selection stage.
Given the candidate network set, the target network for handover is selected by FAHP-based algorithm considering user preferences, QoS requirements and real-time network attributes.
FAHP is proposed as a systemic analysis method~\cite{FAHP} to handle the various characteristics of the decision criteria with a combination of qualitative and quantitative information. \color{black} In our model, FAHP is applied to quantify  the relative importance between pairwise network attributes into the attribute weight vector $\boldsymbol{W}^{\mathrm{S}}_{l}$.  \color{black}

\begin{figure}[htbp!]
    \centering
    \vspace{-0.4cm}
    \includegraphics[width = 0.48\textwidth,height = 0.28\textwidth]{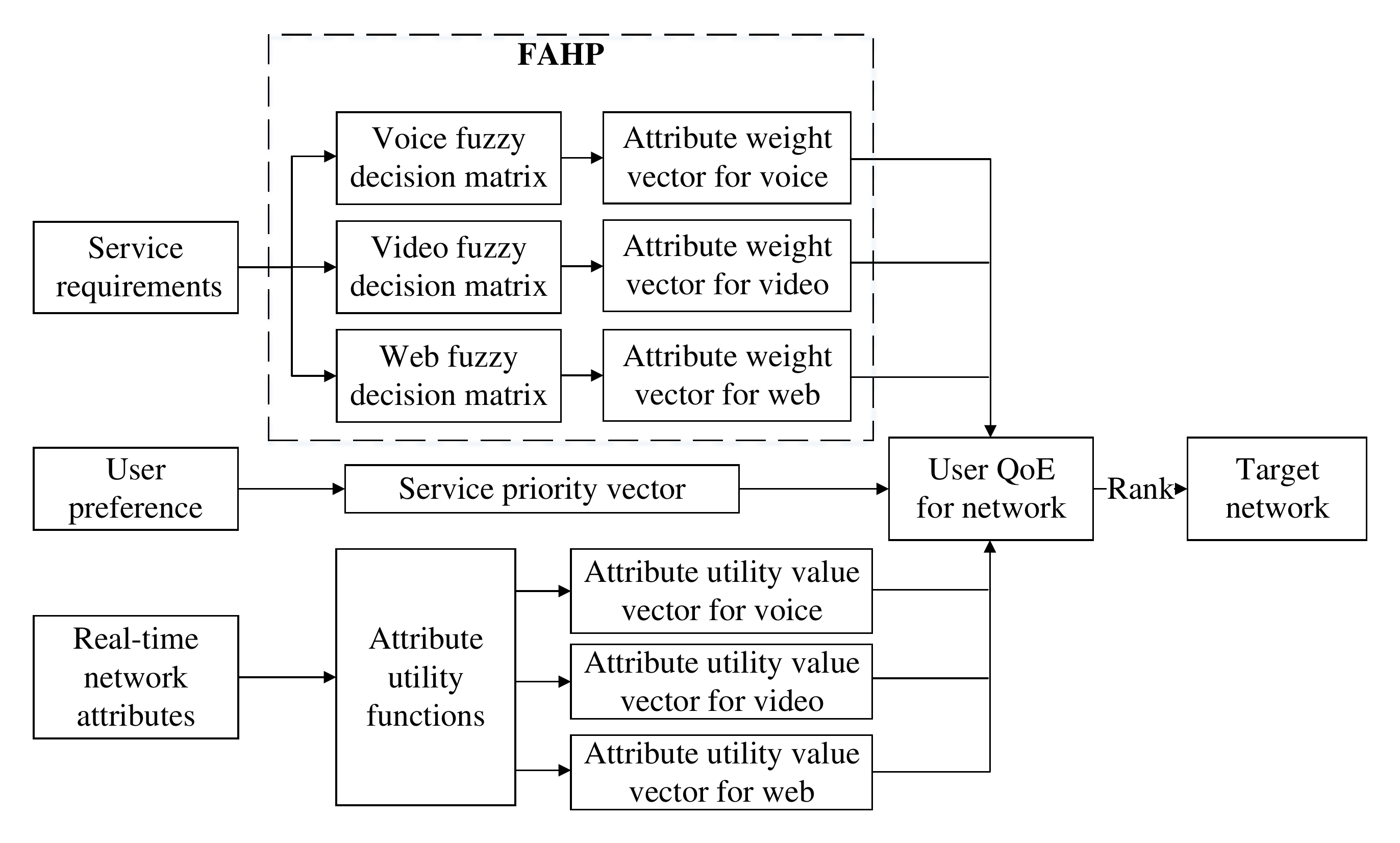}
    \setlength{\abovecaptionskip}{0.cm}
    \vspace{-0.4cm}
    \caption{The framework of FAHP based network selection algorithm.}\label{fig:algorithmframe}
    \vspace{-0.2cm}
\end{figure}
\color{black}
The framework of the FAHP-based network selection algorithm is shown in Fig.~\ref{fig:algorithmframe}.
In the framework, service requirements determine the attribute weights.
User preferences for different services are denoted as service priority vectors specified by users.
The real-time network attributes are normalized by attribute utility functions.
With the above information collected by SDN controller, the target network defined in~(\ref{equ:argmax}) is calculated by the controller. Then, 
MMT executes a seamless handover with SDN control signaling. 
\color{black}

FAHP is applied to calculate the attribute weights $\boldsymbol{W}^{\mathrm{S}}_{l}$.
According to~\cite{select}, the relative importance between the pairwise attribute criteria is indicated by fuzzy numbers. 
With fuzzy numbers, the relative importance of the attribute is extracted as a weight vector. Then, the QoE value is calculated by combining the weight vector, the service priority vector, and the attribute utility value.
\color{black}
In our model, the triangular fuzzy number~(TFN) is introduced to represent the relative importance between pairwise attributes.
The TFN is defined as $\mu = (l,m,u)$, where  $l \le m \le u$, and $l, m, u$ represent the lower limit value, the most favorable value, and the upper limit \color{black} value of the relative importance, respectively. \color{black}
The correspondence of the relative importance between pairwise attributes and TFN value is shown in Table~\ref{tab1}.

\begin{table}[t]
    
    \caption{Importance of TFN value.}
    \begin{center}
        \begin{tabular}{c c c c}
            \hline
            \textbf{No.}&\textbf{Definition}&\textbf{TFN}&\textbf{Reciprocal of TFN} \\
            \hline
            \cline{1-4}
            1 & Equal Importance & (1,1,3) & (1,1,0.33)            \\
            2 & Intermediate Values & (1,2,4) & (0.25,0.5,1)    \\
            3 & Moderate Importance & (1,3,5) & (0.2,0.33,1)    \\
            4 & Intermediate Values & (2,4,6) & (0.17,0.25,0.5)    \\
            5 & Strong Importance & (3,5,7) & (0.14,0.2,0.33)    \\
            6 & Intermediate Values & (4,6,8) & (0.125,0.17,0.25)    \\
            7 & Very Strong Importance & (5,7,9) & (0.11,0.14,0.2)    \\
            8 & Intermediate Values & (6,8,9) & (0.11, 0.125,0.17)    \\
            9 & Extreme Importance & (7,9,9) & (0.11,0.11,0.14)        \\
            \hline
            
        \end{tabular}
        \label{tab1}
    \end{center}
    \vspace{-0.6cm}
\end{table}

Given two different TFNs $\mu_1=(l_1,m_1,u_1)$, $\mu_2=(l_2,m_2,u_2)$,  the summation, multiplication, and reciprocal operations of TFNs are defined as follows:  
\begin{equation}   
    \label{equ2}
    {\mu _1} + {\mu _2}{\rm{ = }}\left( {{l_1} + {l_2},{m_1} + {m_2},{u_1} + {u_2}} \right) ,
    \end{equation}
    \begin{equation}   
    \label{equ3}
    {\mu _1} \otimes {\mu _2}{\rm{ = }}\left( {{l_1} \times {l_2},{m_1} \times {m_2},{u_1} \times {u_2}} \right) ,
    \end{equation}
    \begin{equation}   
    \label{equ4}
    \frac{1}{{{\mu _1}}}{\rm{ = }}\left( {\frac{1}{{{u_1}}},\frac{1}{{{m_1}}},\frac{1}{{{l_1}}}} \right)  .
\end{equation}

  \begin{table*}[t]
    \caption{Service Qos requirements, utility functions and parameters of multiple services.}
    \begin{center}
        \begin{tabular}{c c c c c c c}
            \hline
            \textbf{Service/Attributes}&\textbf{RRS(dBm)}&\textbf{Bandwidth(kbs)}&\textbf{Delay(ms)}&\textbf{Jitter(ms)}&\textbf{Loss Rate(\%)}&\textbf{Cost} \\
            \hline
                    &-85 $ \sim $ -30 &32 $ \sim $ 64&50 $ \sim $ 100& 50 $ \sim $ 80&$ < $ 30&$ < $ 50   \\
            Voice   & $f(x)$ &  $f(x)$ & $g(x)$ & $g(x)$ & $h(x)$ & $h(x)$  \\
                    & $a=0.15,b=-80$ & $a=0.25,b=48$ & $a=0.1,b=75$ & $a=0.185,b=65$ & $g=1/30$ & $g=1/50$  \\
            \hline
                    &-85 $ \sim $ -30 &512 $ \sim $ 5000 &75 $ \sim $ 150&40 $ \sim $ 70&$ < $ 30&$ < $ 50            \\
            Video & $f(x)$ & $f(x)$ & $g(x)$ & $g(x)$ & $h(x)$ & $h(x)$          \\
                  & $a=0.15,b=-80$ & $a=0.003,b=2000$ & $a=0.1,b=112.5$ & $a=0.175,b=55$ & $g=1/30$ & $g=1/50$ \\
            \hline
                &-85 $ \sim $ -30 &128 $ \sim $ 1000&250 $ \sim $ 500&10 $ \sim $ 150&$ < $ 30&$ < $ 50        \\
            Web browsing & $f(x)$ &$ f(x)$ & $g(x)$  & $g(x)$ & $h(x)$ & $h(x)$          \\
            & $a=0.15,b=-80$ & $a=0.01,b=564$ & $a=0.03,b=375$ & $a=0.05,b=80$ & $g=1/30$ & $g=1/50$   \\
            \hline
        \end{tabular}
        \label{ufunc}
    \end{center}
    \vspace{-0.4cm}
\end{table*}

\color{black}
In FAHP algorithm, we take 6 network attributes into account as selection criteria: received signal strength~(RRS), bandwidth, delay, jitter, packet loss rate, and cost.
The number of network attributes is denoted as $N$, and the fuzzy decision matrix   $\boldsymbol{A}^l$ for service $s_l$ is denoted as $\boldsymbol{A}^l = {({\mu _{ij}})_{N \times N}} $. Generally, $\boldsymbol{A}^l$ satisfies $\mu_{ii}=(1,1,3)$ and $\mu_{ji} = 1 / \mu_{ij}$ for $i \ne j$,  $i,j=1\dots N$.
The fuzzy decision matrix $\boldsymbol{A}^l$ is required to hold consistency for the ranking of the importance among criteria. To achieve this, a comprehensive fuzzy value $ F_i$ for attribute $c_i$ is introduced to obtain its weight $ {w}^{\mathrm{S}}_{l,i}$.
$F_i$ is calculated as follows: 
\vspace{-0.2cm}
\begin{equation}   
    \label{equ:Fi}
    {F_i} = \sum\limits_{j = 1}^N {{\mu _{ij}}}  \otimes {\left[ {\sum\limits_{i = 1}^N {\sum\limits_{j = 1}^N {{\mu _{ij}}} } } \right]^{ - 1}} ,
\end{equation}   

The absolute importance of attribute $c_j$ is extracted from the value $V({F_j} \ge {F_i})$, which represents the probability that $F_j$ is larger than $F_i$: 
\vspace{-0.1cm}
\begin{equation}   
\label{equ7}
    V({F_j} \ge {F_i}) = \left\{ \begin{array}{l}
    1\;\;\;\;\;\;\;\;\;\;\;\;\;\;\;\;\;\;\;\;\;\;\;\;\;,\;{m_j} \ge {m_i},\\
    \frac{{({m_j} - {u_j}) - ({m_j} - {l_i})}}{{{l_j} - {u_i}}}\;\;,\;{m_j} \le {m_i},{l_i} \le {u_j},\\
    0\;\;\;\;\;\;\;\;\;\;\;\;\;\;\;\;\;\;\;\;\;\;\;\;\;,\;\mathrm{otherwise}.
    \end{array} \right.   
\end{equation}  

The primary weight value ${w}^{\mathrm{S'}}_{l,j}$ of attribute $c_j$ for service $s_l$ can be defined by (\ref{equ8}) as the minimum probability when $F_j$ is larger than any other $F_i$:
\begin{equation}   
\label{equ8}
    \begin{array}{l}
    {w}^{\mathrm{S'}}_{l,j} = \min V({F_j} \ge {F_i}) = \min V({F_j} \ge {F_1},{F_2}, \ldots ,{F_N}),\\
    \;\;\;\;\;\;\;\;\;\;\;\;\;\;\;\;\;\;\;\;\;\;\;\;\;\;\;\;\;\;\;\;\;\;\;\;\;\;\;\;\;\;\;\;\;\;\;\;\;\;\;\;\;\;j = 1, \ldots ,N.
    \end{array} 
\end{equation}  

Then, attribute weight ${w}^{\mathrm{S}}_{l,j}$ is obtained from ${w}^{\mathrm{S'}}_{l,j}$ as follows:

\vspace{-0.2cm}
\begin{equation}   
\label{equ9}
    {w}^{\mathrm{S}}_{l,j} = \frac{{{w}^{\mathrm{S'}}_{l,j}}}{{\sum\nolimits_{j = 1}^N {{w}^{\mathrm{S'}}_{l,j}} }},{\kern 1pt} {\kern 1pt} {\kern 1pt} {\kern 1pt} {\kern 1pt} {\kern 1pt} j = 1, \ldots ,N,
\end{equation} 
where ${w}^{\mathrm{S}}_{l,j}$ satisfies $\sum\nolimits_{j = 1}^N {{w}^{\mathrm{S}}_{l,j}} {\rm{ = }}1$.
With the above calculation, the final network attribute weight vector for service $s_l$ is denoted as $
\boldsymbol{W}^{\mathrm S}_{l} = [w^{\mathrm S}_{l,1}, w^{\mathrm S}_{l,2},\dots w^{\mathrm S}_{l,N}]$.

\color{black}

When normalizing the network attributes, different utility functions are introduced because the network attributes are classified into benefit attributes and cost attributes.
For attributes with bilateral constraints, the sigmoid utility function is applied for normalization. 
Specifically, $f(x)$ and $g(x)$ are used for benefit attributes and cost attributes, respectively. $f(x)$ and $g(x)$ are presented as follows:

\vspace{-0.1cm}
\begin{equation}   
\label{equ10}
f(x) = \frac{1}{{1 + {e^{ - a(x - b)}}}}  ,
\end{equation}
\begin{equation}   
\label{equ11}
g(x) = 1 - f(x)   ,
\end{equation}
where $a$ and $b$ are constant coefficients determined according to the service requirements for network attributes.
For attributes with unilateral constraints, linear and inverse proportional utility functions are used to adapt to the attribute elasticity. Concretely, $u(x)$ is used for benefit attributes, and $h(x)$ is used for cost attributes. The definitions are presented as follows:
\begin{equation}   
    \label{equ12}
    u(x)=1-g/x ,
\end{equation}
\begin{equation}   
    \label{equ13}
    h(x) = 1 - g \cdot x  ,
\end{equation}
where $g$ is a constant coefficient determined by the service requirements.

\begin{figure}[t]
    \centering
    \includegraphics[width=0.45\textwidth,height = 0.3\textwidth]{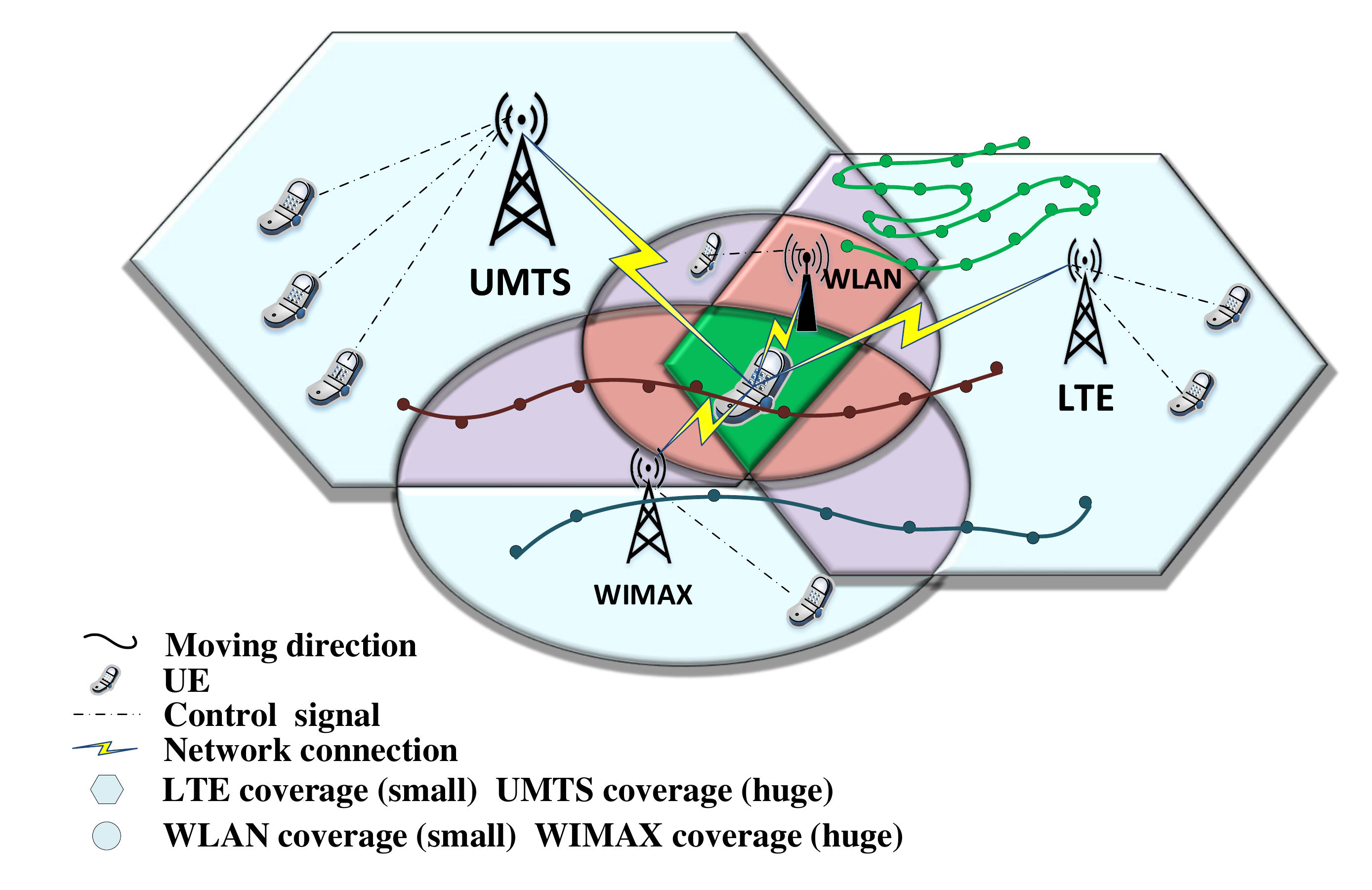}
    \vspace{-0.4cm}
    \caption{Handover scenario and moving patterns in HetNets.}\label{fig:moving_scenario}
    \vspace{-0.4cm}
\end{figure}

\color{black}

The raw network attribute values collected by SDN controller are constructed in matrix form as~(\ref{equ:real-time attrib}).
The utility functions and their coefficients are diverse among different services, and are shown in Table~\ref{ufunc}. 
Then, the normalized attribute utility $\boldsymbol{U}^i_l$ of service $s_l$ is calculated from $\boldsymbol{A}_{t}$ with the utility functions. 
The QoE metric of the network combines user-determined service priority $\boldsymbol{P}_k$, service-determined attribute weight $\boldsymbol{W}^{\mathrm{S}}_{l}$ and real-time network attribute utility value $\boldsymbol{U}_l^i$. 
 Given a candidate network set $\mathcal{R}$, the QoE $S_i$ of network $r_i$ is defined as~(\ref{equ:QoE}).
The target network is selected with the highest QoE according to~(\ref{equ:argmax}) to execute the handover process. 
Given the target network, 
the handover process would be triggered only if $S_t/S_i > \delta$, where $S_t$, $S_i$, and $\delta$ represent the target network's QoE value, the originally connected network's QoE value and the handover threshold, respectively.

\color{black}

Generally, when RSS undergoes drastic fluctuations due to network variance or small-scale movement, the MMT may require frequent handovers between two networks, which causes ping-pong effect that wastes network resources. 
Fig.~\ref{fig:moving_scenario} shows several moving patterns during handover. In Fig.~\ref{fig:moving_scenario}, we can see that when the user's moving trajectory is like the green line, the probability that ping-pong effect occurs will increase considerably.
With SDN control plane storing the user's historical location coordinates, a ping-pong moving pattern can be recognized. In our method, according to the AP's coverage radius, we mark a potential ping-pong moving pattern in which the standard deviation $\sigma$ of user's $N_{in}$ locations has not surpassed an indoor spatial distance. We removed this pattern from handover trigger conditions, thus avoiding ping-pong effect. 
After a handover condition is triggered, the MMT's handover execution towards the target network is instructed by SDN controller using MPTCP.

    \vspace{-0.1cm}
    \subsection{MPTCP-Based Seamless Handover Execution} 
    \color{black}
    In this subsection, we propose a signaling process for MPTCP-based seamless handover execution in SDN.
    We will first illustrate the mechanism and path management strategies of MPTCP. Next, we propose a signaling flow chart for MPTCP-based seamless handover.
\begin{figure}[t]
    \centering
    \includegraphics[width=0.45\textwidth, height=0.23\textwidth
    ]{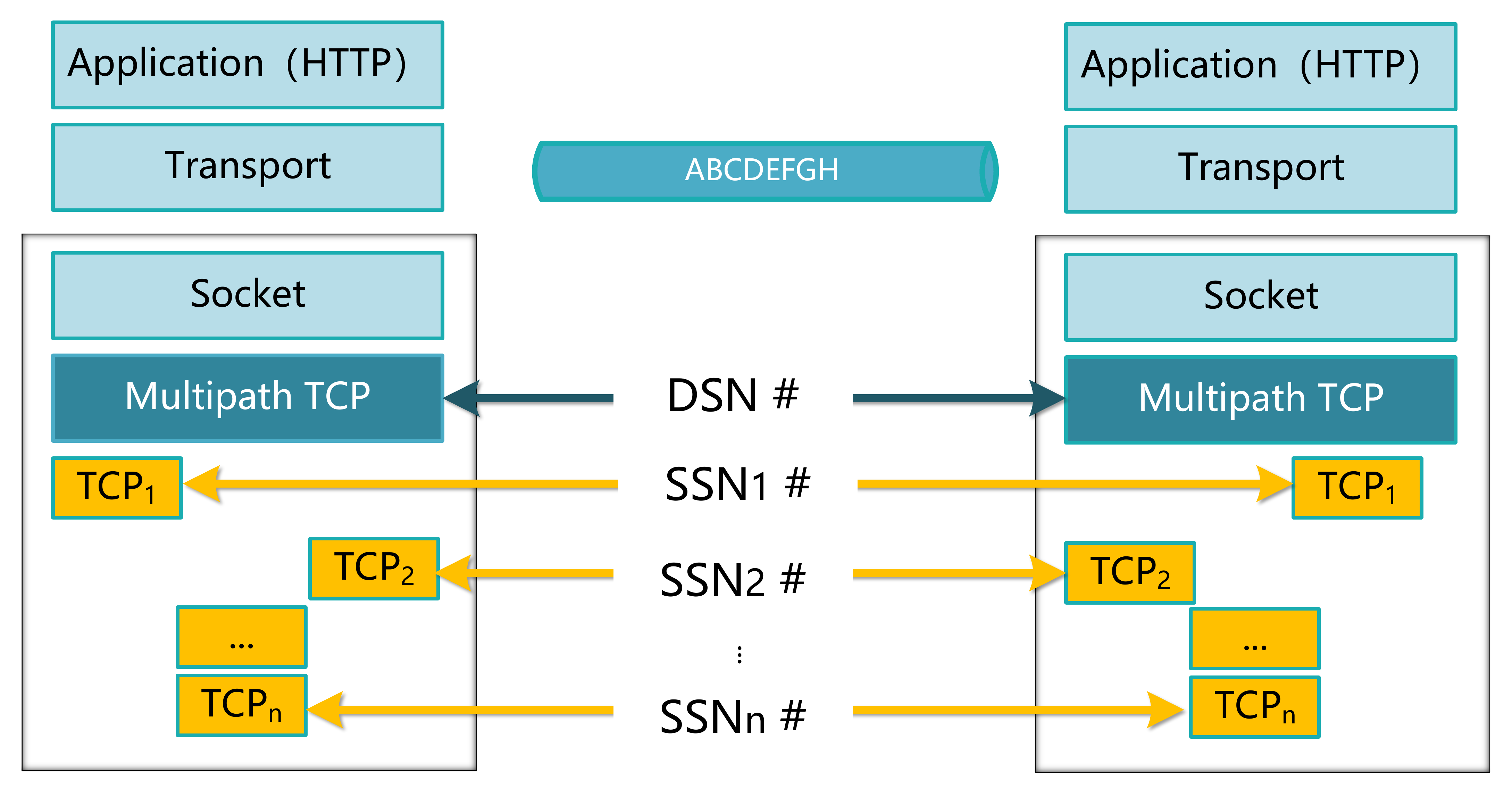}
    \caption{Architecture of MPTCP.}\label{fig:MPTCP architecture}
    \vspace{-0.5cm}
\end{figure}
    \color{black}

    \subsubsection{Mechanism of MPTCP} 
    For one TCP/IP session, there are usually multiple paths available for data transmission, such as routing paths through WLAN and LTE networks.
    The simultaneous use of the multiple paths for one TCP/IP session would improve user experience with a higher throughput and resilience to network failure. 
        Proposed by IETF, MPTCP is an extension of TCP with the capability to simultaneously use multiple paths between peers~\cite{rfc6824}.
        MPTCP can establish multiple subflow connections across disjoint paths and communicate through the subflows.
    Evolving from the conventional TCP mechanism,
    MPTCP offers the same type of functions for application layer as TCP, such as the reliable byte stream transmission, etc.
    To keep the packets from different paths in order, in the frame structure, MPTCP configures data sequence number~(DSN), subflow sequence number~(SSN) and data sequence mapping~(DSM). 
    DSN indicates the order of all packets in the MPTCP session, while SSN denotes the packet order in corresponding subflow. DSM represents the maps from SSN to DSN. 
    DSM is fixed once it is specified, and DSM is transmitted along with the packet segment~\cite{MPTCP}. 
    A simplified  MPTCP architecture is shown as Fig.~\ref{fig:MPTCP architecture}.

    The advantage of MPTCP is that it can simultaneously use multiple subflows through different paths between peers. MPTCP can split the traffic flow of one MPTCP session into many subflows.
    For each subflow's data communication, MPTCP can establish one independent TCP sub-session.
    At the receiving end, the received data packages from different subflows are reassembled into a single flow according to DSN. With the simultaneous use of multiple subflows, MPTCP can make full use of network resources, thus improving user experience with higher throughputs and reliable resilience to network failure.

    \begin{figure}[t]
        \centering
        \includegraphics[width=0.3\textwidth, height=0.2\textwidth
        ]{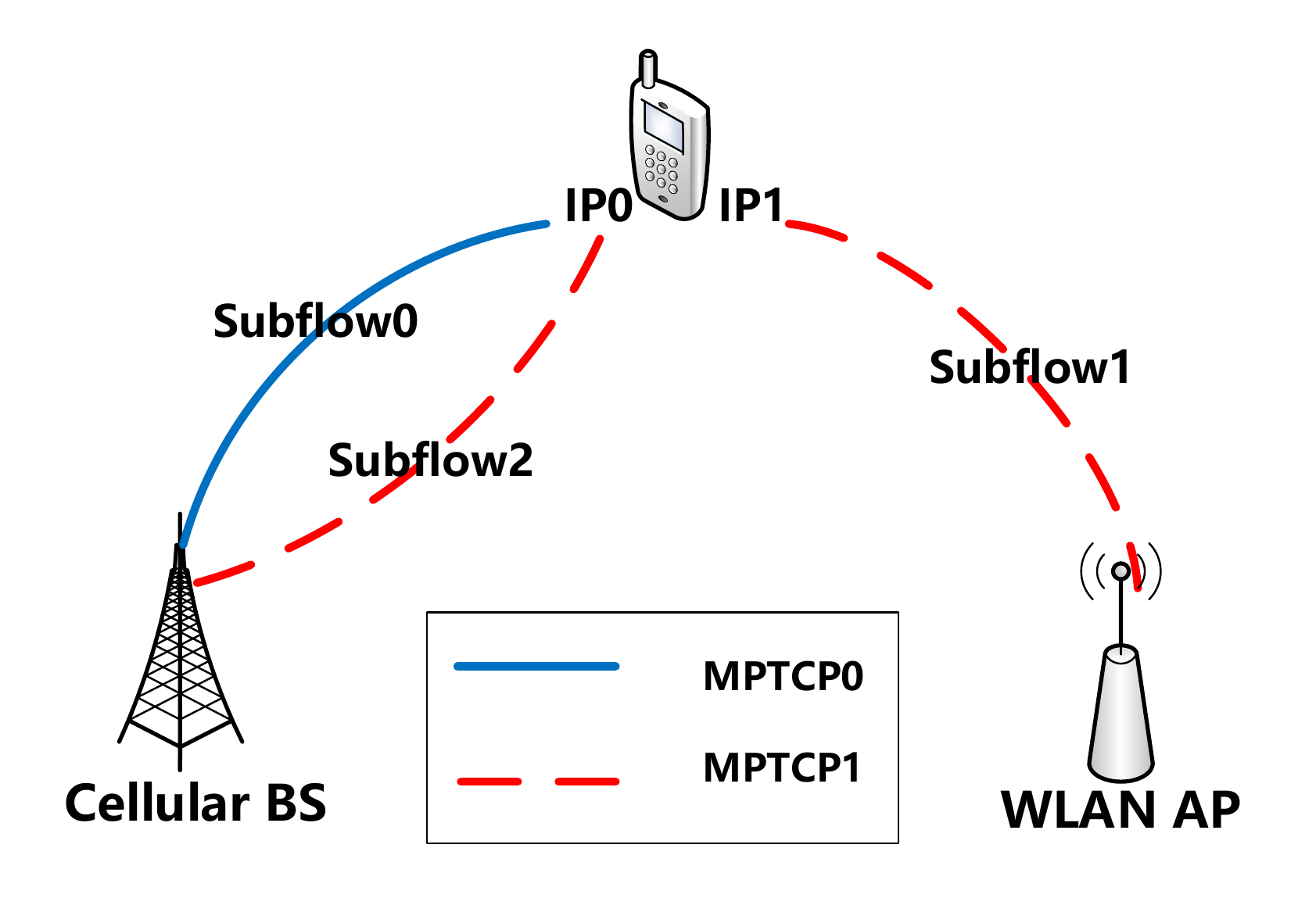}
        \caption{\color{black} Original and new MPTCP sessions in mobility caused handover process. The existing session MPTCP0 originally established one subflow for limited network coverage area. While the new session MPTCP1 can establish two subflows through cellular base station and WLAN AP, as the MMT moves into the overlapped area of cellular BS and WLAN AP. \color{black}}\label{fig:Old and New MPTCP Sessions}
        \vspace{-1em}
    \end{figure}
    
    \subsubsection{Path Management Strategies of MPTCP}
    MPTCP has been implemented on Linux Kernel with several optional strategies to establish new subflows and schedule the available subflows. 
    MPTCP declares a modular structure called path-manager to establish new subflows. With a specified path-manager, the host will establish new subflows according to the configuration.
    \color{black} MPTCP also declares a modular infrastructure called scheduler, which schedules the use of subflows for packages' transmission.
    With different path-manager and scheduler strategies, MPTCP can transmit data in an effective or reliable manner according to the configuration.
    A brief introduction of MPTCP path-manager and scheduler strategies is presented as follows~\cite{three_modes}:
    \color{black}
    \begin{itemize}
        \item 
        Fullmesh: 
        The fullmesh path-manager mode creates a full mesh of subflows among all available IPs between peers.
        These subflows can use different paths through HetNets to transmit data.
        \item
        \color{black}Redundant: 
        The redundant scheduler mode configures MPTCP to simultaneously transmit data through all available subflows in a redundant way. This scheduler mode fits the situation where the user requires the lowest potential latency communication by sacrificing bandwidth utilization.
        \color{black}
        \item 
        Default: In default scheduler mode, MPTCP transmits data on the subflow with the lowest round trip time~(RTT) until its congestion window is full. Then, MPTCP starts to transmit data on the subflows with the next higher RTT. 
        This scheduler mode fits the situation when the user requires achieving  higher throughputs by using subflows with the best channel condition.
    \end{itemize}
    \color{black} 
    
    Based on the above definition, during handover
    we set the path-manager in fullmesh mode, so as to use subflows through multiple paths from HetNets. Then, we configure the default scheduler mode to achieve higher throughputs during the handover process. 
    Using these configurations, MPTCP still cannot decide when to switch path-manager mode nor scheduler mode.
    To solve this problem, we propose a signaling process using SDN controller's command to instruct the MMT establishing a new MPTCP subflow for forthcoming handover.
    \color{black}

    \begin{figure}[t]
        \centering
        \includegraphics[width=0.4\textwidth, height=0.33\textwidth    
        ]{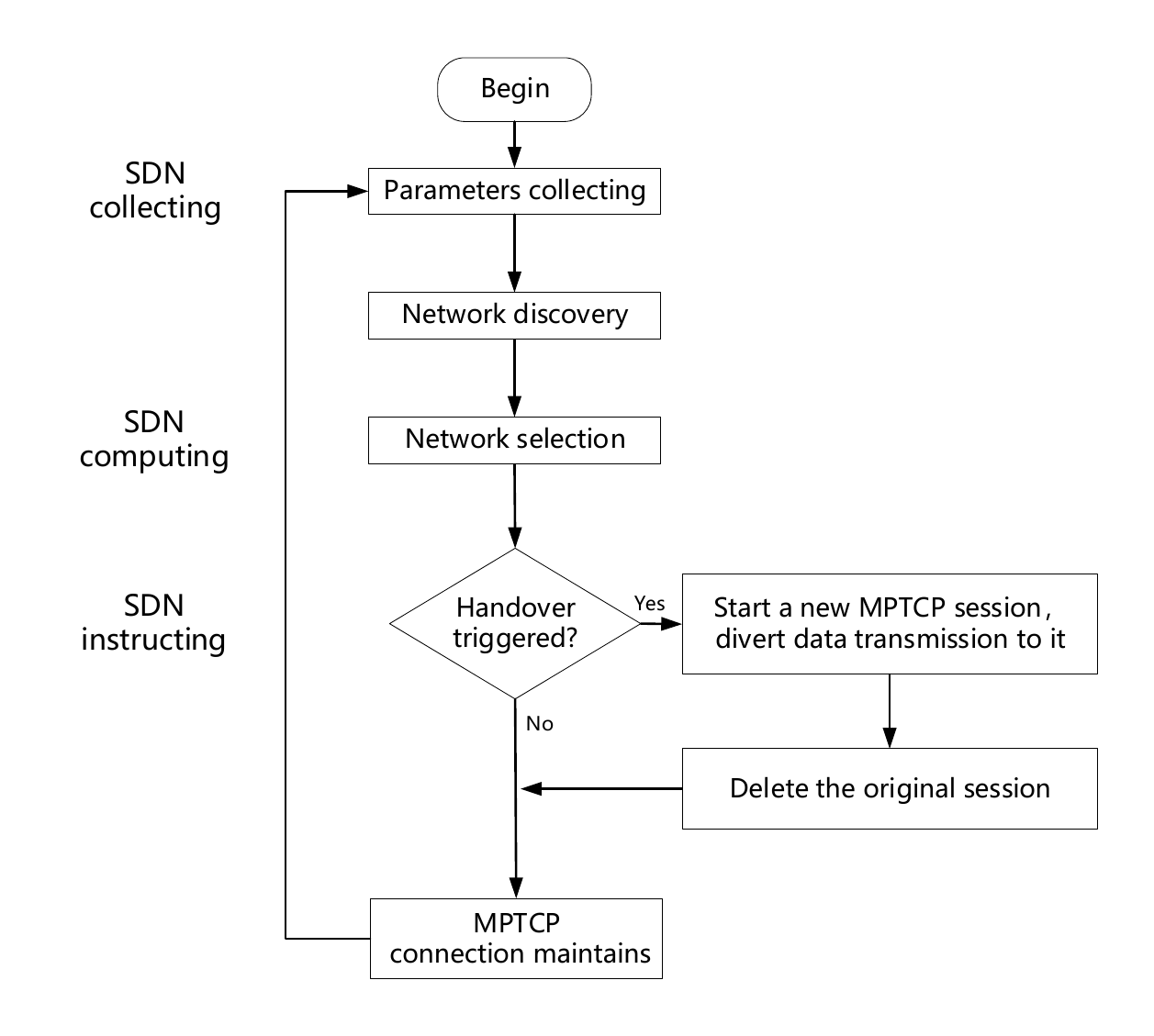}
        \caption{Process of proposed seamless handover method. }\label{fig:signaling flow chart}
        \vspace{-0.6cm}
    \end{figure}

    \vspace{-0.3cm}
    \subsubsection{Seamless Handover Execution}
    
    In order to guarantee service continuity, the handover execution process is completed with the cooperation of SDN controller and the MMT. SDN controller periodically collects the network attributes, runs the network selection algorithm, and instructs the network handover of MMT. As shown in Fig.~\ref{fig:Old and New MPTCP Sessions}, when the original MPTCP session, MPTCP0, fades and the new available networks’ performance improves (in this situation the handover is triggered), SDN controller will send a handover command to initiate a new MPTCP session, MPTCP1, and \color{black} gradually \color{black} divert the service data transmission from MPTCP0 to MPTCP1. 
    Note that under fullmesh mode, MPTCP1 creates multiple subflows through all available HetNets. To save network resources, MPTCP0 is terminated after MPTCP1 is established. The entire handover process is shown in Fig.~\ref{fig:signaling flow chart}.
 
    \begin{figure}[t]
        \centering
        \includegraphics[width=0.5\textwidth, height=0.4\textwidth
        ]{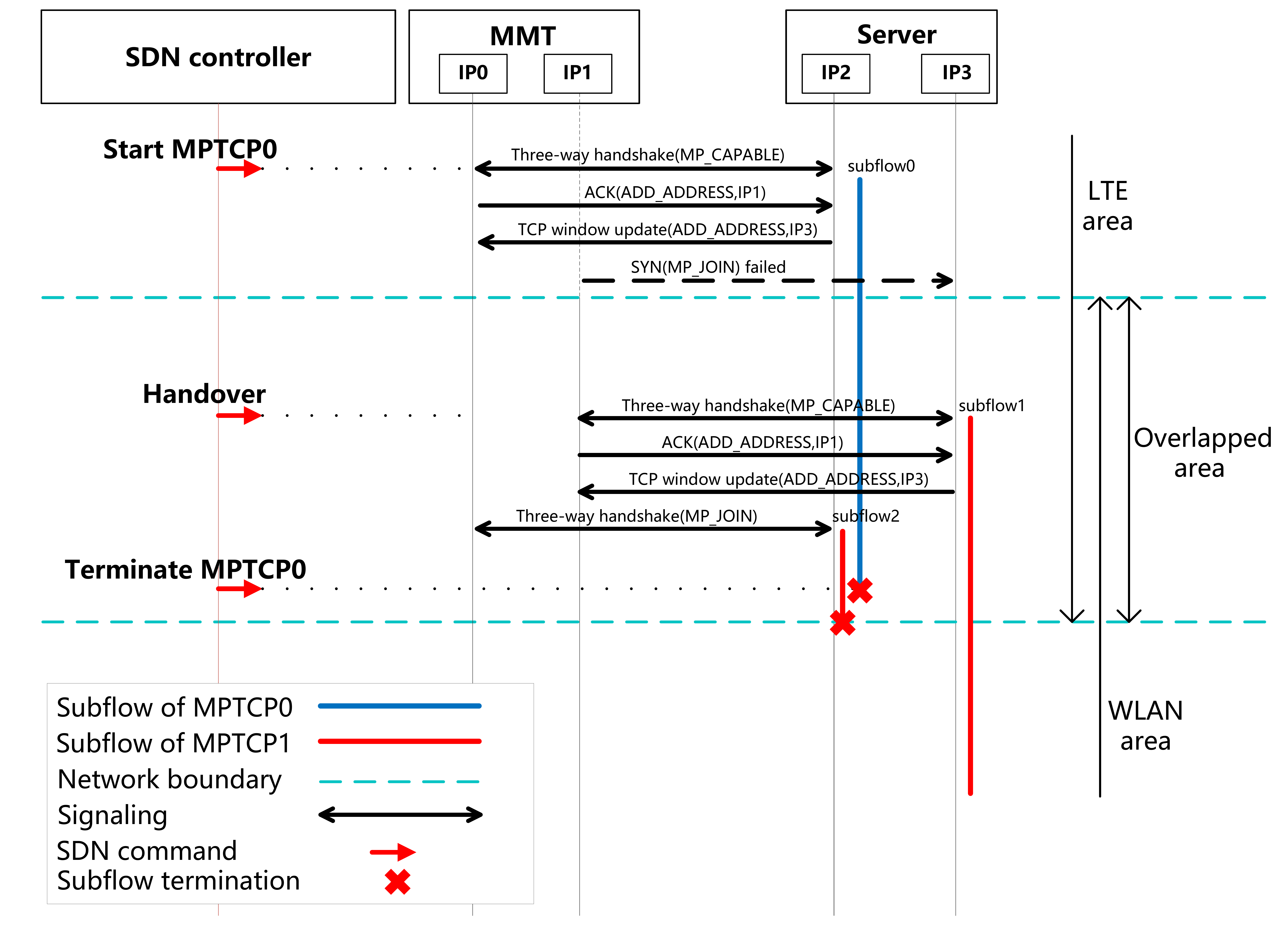}
        \caption{Signaling flow chart for MPTCP based seamless handover.}\label{fig:message sequence chart}
        \vspace{-0.6cm}
    \end{figure}
    \color{black}
    To further elaborate on the handover process, as an illustration, we consider a scenario such as Fig.~\ref{fig:Old and New MPTCP Sessions}, in which the MMT moves from outdoors to indoors, receiving service data from the service provider's remote server, and the MMT tends to handover from LTE to WLAN.
    In this process, both the MMT and the server are equipped with two interfaces and can communicate through MPTCP.
    The signaling flow chart of the seamless handover execution is presented in Fig.~\ref{fig:message sequence chart} and elaborated as follows:
  \begin{itemize}
        \item 
        SDN controller periodically collects network attributes and runs the network selection algorithm to determine the target network and whether to trigger handover. In the outdoor area where only LTE covers, MPTCP0 is initialized for data transmission through LTE, in which subflow0 between IP0 and IP2 is established with a three-way handshake procedure. Then, to exchange the information of extra IP addresses,
        the MMT enables ADD\_ADDRESS packets and sends them to the server.
        After that, the MMT attempts to create another subflow by sending an SYN packet from IP1 to IP3, but fails because IP1 has no access to WLAN.
        \item 
        When the MMT moves into the overlapped area between LTE and WLAN and the handover condition is triggered, SDN controller sends a handover command to initialize MPTCP1 \color{black}in fullmesh path-manager mode \color{black} and divert the service data transmission from MPTCP0 to MPTCP1. 
        Unlike MPTCP0, MPTCP1 succeeds in establishing a second subflow because the MMT can currently access to both LTE and WLAN. The subflow transmission through WLAN is marked as subflow1, and the other subflow through LTE is marked as subflow2. 
        \item 
        To save network resources, MPTCP0 will be terminated when data transmission over it finishes. When the MMT moves into the indoor area where only WLAN covers, subflow1 of MPTCP1 fades away as LTE signal decays. However, subflow2 of MPTCP1 maintains data transmission through WLAN, thus realizing seamless handover and guaranteeing service continuity. \color{black}
    \end{itemize}

\section{Simulation and Performance Analysis}

\renewcommand\arraystretch{1.2}
\begin{table}[t]
\vspace{-0.4cm}
\centering
    \caption{Simulation Parameters.}
        \begin{tabular}{|c|c |c|c| }         
        \hline 
        \textbf{Parameter}&\textbf{Value}&\textbf{Parameter}&\textbf{Value}\\
        \hline
        $L$                & 3      & $N$               & 6           \\
        \hline
        $s_1$              & voice  & $s_2$         & video         \\
        \hline
        $s_3$              & web browsing & $M$      & 4    \\
        \hline
        ${P}_{\rm{tB,UMTS}}$ & 53 W & ${P}_{\rm{tB,LTE}}$ & 43 W       \\
        \hline
        ${P}_{\rm{tB,WiMAX}}$ & 30 W  & ${P}_{\rm{tW,WLAN}}$ & 23 W     \\
        \hline
        $N_{in}$        &  4   & $N_{s}$           & 1           \\
        \hline
        $W$                & 64   & $\lambda$         & 0.01        \\
        \hline
        $H$                &  30 s & $N_{tr}$         & 6800        \\ 
        \hline
        $f_W$             & 2400 MHz &  $\delta$          & 1.08    \\
        \hline
         $\sigma$            & 12.9  &  $\boldsymbol {P}_{\rm 1}$ & $[0.1\; 0.3 \; 0.6]$    \\
        \hline
        $\boldsymbol {P}_{\rm 2}$&$[0.1 \; 0.6\; 0.3]$ & $\boldsymbol {P}_{\rm 3}$ & $[0.6 \; 0.3\; 0.1]$ \\
        \hline
        \color{black}$\sigma B$      & \color{black} 2.296 &  \color{black}$\sigma W$    &\color{black} 2.303 \\
        \hline
        \end{tabular}
    \label{tab:simu_peremeter}
    \vspace{-0.2cm}
\end{table}

For our simulation, the mobile user's location data that the ESN uses to train and infer are obtained from \emph{CRAWDAD} of \emph{Human mobility data}~\cite{KAISTdataset}.
The user location data are measured by GPS receivers every 30 seconds.
Here, the SDHetNets consist of $M=4$ types of networks. \color{black} $U=90$ users are considered moving casually, and their location coordinates are uploaded every 30~s. \color{black} Each type of network access points are uniformly distributed according to their coverage radius. The detailed parameters are listed in Table~\ref{tab:simu_peremeter}. All statistical results are calculated based on 10000 independent runs. The accuracy of the prediction experiment is measured by normalized root mean square error~(NRMSE)~\cite{conceptor_esn}.

\color{black}
For location prediction simulation, we divided the dataset into cross-region moving pattern (move exceeding an area of 400 m in 1 hour) and ping-pong moving pattern (movement within an area of 50m in 1 hour).
We structured the training data by extracting the data evenly from the cross-region and ping-pong moving pattern data, so as to accelerate the ESN extracting the users' transfer relationship in different locations. 
The ratio of the training set and cross-validation set to the test set is 6:2:2.
The performance of the proposed ESN model is compared with Long Short-Term Memory~(LSTM)~\cite{LSTM} and Gated Recurrent Unit~(GRU)~\cite{GRU}.\color{black}

For network selection simulation, we apply the fuzzy decision matrices as shown in Table~\ref{tab:decmtrix}. 
RSS is calculated from the predicted locations following the transmission model~(\ref{equ:hata231}),~(\ref{equ:wlanloss}), and the other attributes are set to vary randomly following Table~\ref{ufunc}.
For comparison purposes, we simulate:
a) RSS-based handover method, 
b) fuzzy multiple criteria group decision making~(Fuzzy MCGDM) algorithm~\cite{MCGDM}, 
c) fuzzy technique for order preference by similarity to the ideal
situation (FTOPSIS)~\cite{FTOPSIS},  
d) the proposed FAHP-based algorithm, and
e) the proposed FAHP-based algorithm with uniform  location errors of  10$\sim$20 m (FAHP with errors). 

For handover execution simulation, the proposed method is realized and evaluated in Mininet-WiFi~\cite{kiran} by programming on the SDN controller.
The handover performance is analyzed by data throughputs using Wireshark. 
The throughput result is compared with the conventional handover method using a single interface. 

\color{black}

\begin{table}[b]
    \vspace{-0.6cm}
    \centering
    \caption{Prediction NRMSE of ESN and LSTM as number of neurons varies. 
    }
    \begin{tabular}{|c|c|c|c|c|c|c|}
    \hline
    \multicolumn{6}{|c|}{ESN}  \\
    \hline 
        $W$ & 16 & 32 & 48 & 64 & 80 \\
    \hline
        NRMSE & 0.1456 & 0.1104 & 0.0999 & 0.0949 & 0.0958 \\
    \hline 
    \multicolumn{3}{|c|}{Convergence time} & \multicolumn{3}{|c|}{1 regression step} \\
    \hline
    \multicolumn{6}{|c|}{LSTM}\\
    \hline
    \multicolumn{3}{|c|}{Number of hidden units} & 32 & 64 & 128           \\
    \hline
    \multicolumn{3}{|c|}{NRMSE} & 0.1427 & 0.0813 & 0.0783 \\ 
    \hline 
    \multicolumn{3}{|c|}{\color{black}Convergence time} & \multicolumn{3}{|c|}{\color{black}10 k  $\sim$ 20 k iteration epoches} \\
    \hline
    \multicolumn{6}{|c|}{\color{black}GRU}\\
    \hline
    \multicolumn{3}{|c|}{Number of hidden units}     & \color{black}64 & \color{black}128 & \color{black}256 \\
    \hline
    \multicolumn{3}{|c|}{NRMSE}  & \color{black}0.2765 & \color{black}0.1041 & \color{black}0.0768 \\
    \hline
    \multicolumn{3}{|c|}{Convergence time} & \multicolumn{3}{|c|}{\color{black} 10 k $\sim$ 15 k iteration epoches} \\
    \hline
    \end{tabular}
    \label{tab:nrmse}
\end{table}

\begin{figure}[t]
    \centering
    \vspace{-0.6cm}
    \includegraphics[width=0.45\textwidth, height=0.31\textwidth
    ]{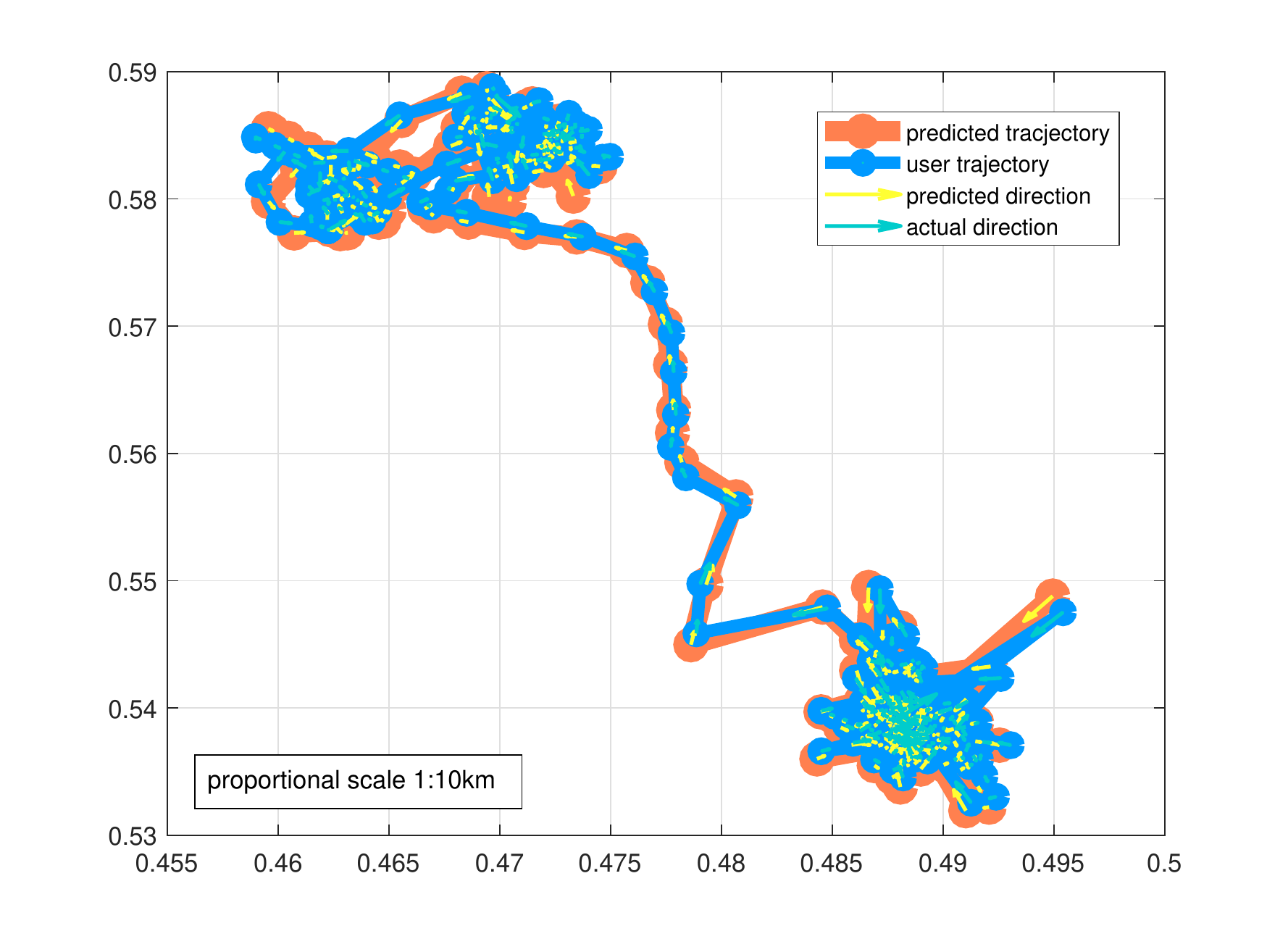}
    \caption{Predicted user trajectory by ESN and actual user trajectory in KAIST. The ESN parameters are set as Table~\ref{tab:simu_peremeter}. }\label{fig:location_pre}
    \vspace{-0.7cm}
\end{figure}

Table~\ref{tab:nrmse} and Fig.~\ref{fig:location_pre}  evaluate the accuracy of ESN predicting the user's future location. 
First, in Fig.~\ref{fig:location_pre}, 
we show how ESN can predict the user's location in a two-dimensional map of KAIST.
In Fig.~\ref{fig:location_pre}, we can see that the user's future trajectory is effectively predicted via ESN perceiving users' moving patterns. 
\color{black}Next, in Table~\ref{tab:nrmse},  we show how NRMSE of ESN's prediction changes as the number of neurons varies and compare ESN with LSTM and GRU, which obtain prominent performances on sequence's long-term feature processing. \color{black}
In Table~\ref{tab:nrmse}, we can see that, as the number of neurons increases, the NRMSE of the proposed ESN model decreases when $W\leq64$, while increases when $W>64$. 
This is because
as the number of neurons increases, the memory capacity of the ESN reservoir increases, and hence the learning ability of ESN improves.
Moreover, as the ESN model is trained from limited data, the prediction error increases because the oversized model causes overfitting.
In Table~\ref{tab:nrmse}, we can also see that 
the ESN can be trained with less convergence time than LSTM and GRU. The prediction NRMSE of ESN gets close to the NRMSE of LSTM and GRU. \color{black} GRU has a relatively lower  convergence time while requiring more model complexity than LSTM for omitting one gate~\cite{GRU}. \color{black} 
Clearly, the proposed ESN model provides a reliable location prediction with small training samples in a faster learning manner than the deep learning models LSTM and GRU, which is suitable for distributed network controller deployment.

\begin{table*}[t]
    \vspace{-0.3cm}
    \caption{ Fuzzy decision matrix of voice, video and web service~\cite{select}.}
    \vspace{-0.6cm}
    \begin{center}
        \setlength{\tabcolsep}{3mm}{
        \begin{tabular}{c c c c c c c c}
            \textbf{Voice Matrix}&&&&&&  \\
            \toprule[2pt]
            &\textbf{RSS}&\textbf{Bandwidth}&\textbf{Delay}&\textbf{Jitter}&\textbf{Loss Rate}&\textbf{Cost}& \textbf{Weight} \\
            \hline
            RSS & (1,1,3) & (3,5,7) & (0.2,0.33,1)& (1,3,5)& (1,3,5)& (0.25,0.5,1)& 0.1776   \\
            Bandwidth & (0.14,0.2,0.33) & (1,1,3) & (0.25,0.5,1)& (0.17,0.25,0.5)& (1,2,4)& (0.125,0.17,0.25)& 0.1115  \\
            Delay & (1,3,5) & (1,2,4) & (1,1,3)& (0.2,0.33,1)& (1,3,5)& (0.14,0.2,0.33)& 0.1605  \\
            Jitter & (0.2,0.33,1) & (2,4,6) & (1,3,5)& (1,1,3)& (3,5,7)& (0.2,0.33,1)& 0.1834  \\
            Loss Rate &(0.2,0.33,1)&(0.25,0.5,1)&(0.2,0.33,1)& (0.14,0.2,0.33)& (1,1,3)& (0.11,0.14,2)& 0.0986  \\
            Cost & (1,2,4)& (4,6,8) & (3,5,7) & (1,3,5)& (5,7,9)& (1,1,3)& 0.2683  \\
            \hline

            \textbf{Video Matrix}&&&&&&  \\ 
            \toprule[2pt]
            &\textbf{RSS}&\textbf{Bandwidth}&\textbf{Delay}&\textbf{Jitter}&\textbf{Loss Rate}&\textbf{Cost} & \textbf{Weight} \\
            \hline
            RSS & (1,1,3) & (0.2,0.33,1) & (3,5,7)& (5,7,9)& (1,3,5)& (1,3,5)& 0.2098   \\
            Bandwidth & (1,3,5) & (1,1,3) & (1,2,4)& (0.17,0.25,0.5)& (4,6,8)& (4,6,8)& 0.2033  \\
            Delay & (0.14,0.2,0.33)  & (0.25,0.5,1) & (1,1,3)& (0.14,0.2,0.33)& (1,3,5)& (1,3,5)& 0.1330  \\
            Jitter & (0.14,0.2,0.33) & (2,4,6) & (3,5,7)& (1,1,3)& (5,7,9)& (5,7,9)& 0.2345  \\
            Loss Rate &(0.2,0.33,1)&(0.125,0.17,0.25)&(0.2,0.33,1)& (0.11,0.14,0.2)& (1,1,3)& (1,1,3)&  0.1097 \\
            Cost & (0.2,0.33,1)&(0.125,0.17,0.25)& (0.2,0.33,1)& (0.11,0.14,0.2)& (1,1,3)& (1,1,3)& 0.1097  \\
            \hline

            \textbf{Web Matrix}&&&&&&  \\
            \toprule[2pt]
            &\textbf{RSS}&\textbf{Bandwidth}&\textbf{Delay}&\textbf{Jitter}&\textbf{Loss Rate}&\textbf{Cost} & \textbf{Weight} \\
            \hline
            RSS & (1,1,3) & (0.2,0.33,1) & (3,5,7)& (3,5,7)& (0.2,0.33,1)& (0.14,0.2,0.33)& 0.1609   \\
            Bandwidth & (1,3,5) & (1,1,3) & (3,5,7)& (4,6,8)&(0.25,0.5,1) & (1,2,4) & 0.2084  \\
            Delay & (0.14,0.2,0.33)  & (0.14,0.2,0.33) & (1,1,3)& (1,2,4)& (0.125,0.17,0.25)& (0.17,0.25,0.5) & 0.1076 \\
            Jitter & (0.14,0.2,0.33) &  (0.125,0.17,0.25)& (0.25,0.5,1)& (1,1,3)& (0.11,0.14,0.2)& (0.14,0.2,0.33)& 0.0897  \\
            Loss Rate &(1,3,5)&(1,2,4)&(4,6,8)& (5,7,9)& (1,1,3)& (1,3,5) & 0.2428 \\
            Cost & (3,5,7)&(0.25,0.5,1)& (2,4,6)& (3,5,7)& (0.2,0.33,1)& (1,1,3) & 0.1905 \\
            \hline
        \end{tabular}
    }
        \label{tab:decmtrix}
    \end{center}
    \vspace{-0.4cm}
\end{table*}

\begin{figure}[htbp!]
    \centering
    \setlength{\abovedisplayskip}{0.pt}
    \vspace{-0.4cm}
    \includegraphics[width=0.5\textwidth, height=0.34\textwidth
    ]{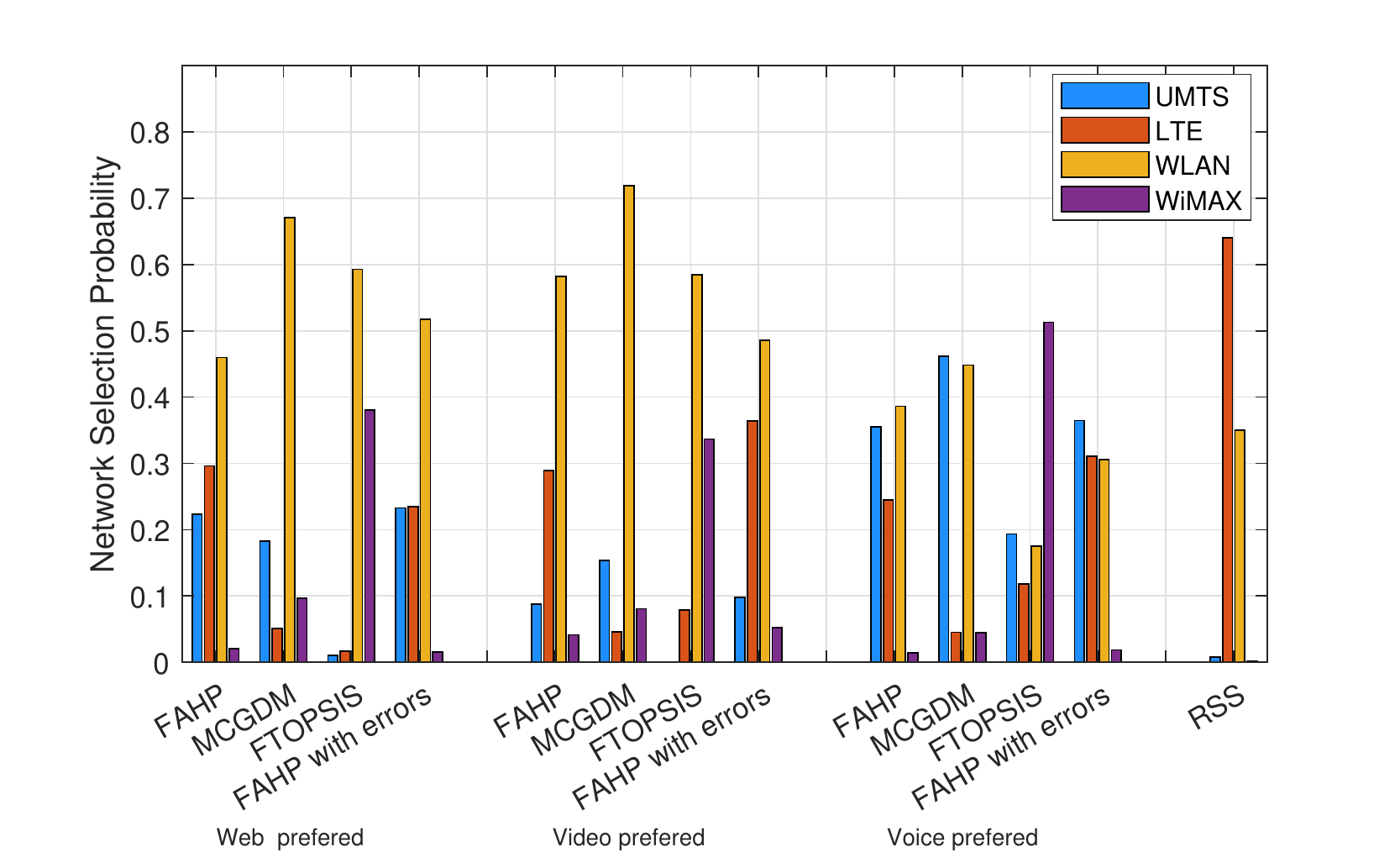}
    \caption{Selected probability for networks with different service priority.
    Service priority vectors, utility functions, and attribute weights are set according to Table~\ref{tab:simu_peremeter}, Table~\ref{ufunc} and Table~\ref{tab:decmtrix}, respectively. For each type of network, network access points are placed geographical evenly 
    with interval that is 35 km, 1 km, 500 m and 100 m for UMTS, LTE, WLAN and WiMAX, respectively~\cite{cellcover}.     
    }\label{fig:selectedprob}
    \setlength{\belowdisplayskip}{3.pt}
    \vspace{-0.6cm}
\end{figure}


Fig.~\ref{fig:selectedprob} shows the statistical probability of each type of network to be selected with different user preferences in 10000 runs.
MCGDM is a decision-making algorithm that can select the appropriate network with adequate consideration of user preferences.
\color{black}
 FTOPSIS algorithm combines fuzzy theory with TOPSIS rank method.
From Fig.~\ref{fig:selectedprob}, we can see that
both the FAHP and the MCGDM algorithms adapt to diverse user preferences better than FTOPSIS. 
That is, when web or video services are preferred, WLAN is more likely to be selected for its high bandwidth and low packet loss rate, while when voice services are preferred, UMTS is selected more frequently for low delay and low jitter.
FTOPSIS is inclined to WLAN but not UMTS, this is because FTOPSIS uses the steady ranking method to make decisions.
In Fig.~\ref{fig:selectedprob}, we can also see that
compared to MCGDM algorithm, FAHP has a more balanced probability of selecting each type of network, which is beneficial for reducing the access load. Moreover, FAHP is more inclined to select LTE, which is more consistent with actual network deployment.
This is because the proposed FAHP-based algorithm combines information from both users and networks for network selection. The introduced fuzzy numbers and sigmoid utility functions enhance the algorithm's adaptability for attribute variation.
To verify the proposed algorithm's robustness against localization and prediction errors, we simulated the algorithm with inaccurate location data labeled ``FAHP with errors". In Fig.~\ref{fig:selectedprob}, we can also see that FAHP maintains a relatively steady selection tendency when meeting errors on locations.
\color{black}

\begin{figure}[t]
    \centering
    \vspace{-0.2cm}
    \includegraphics[width=0.45\textwidth,height=0.33  \textwidth
    ]{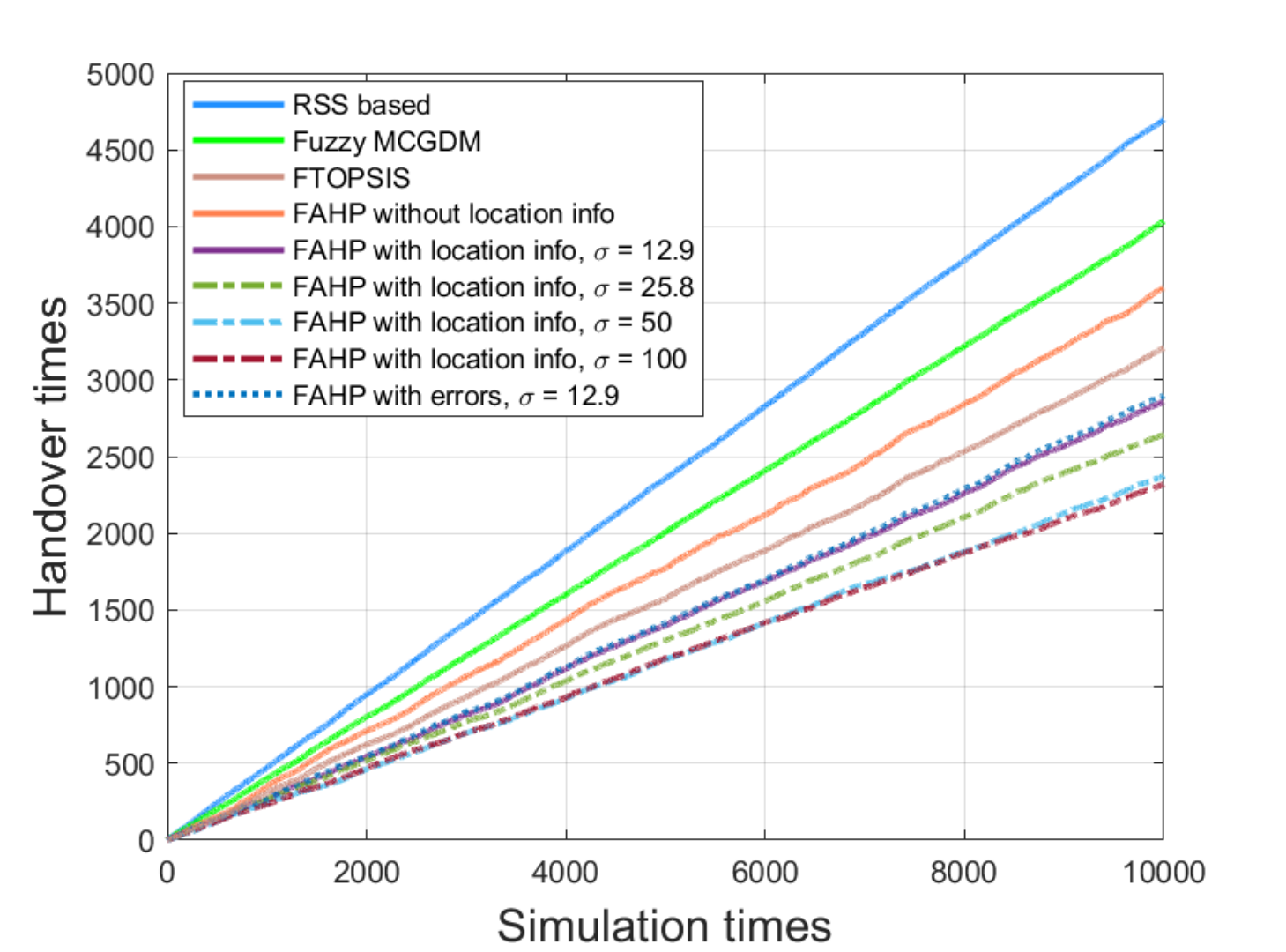}
    \caption{Handover times of different algorithms as simulation times and $\sigma$ vary. The experiment  runs in web service preferred scenario. The threshold parameters are set as Table~\ref{tab:simu_peremeter}. }\label{fig:handovernumber}
    \vspace{-0.4cm}
\end{figure}

\color{black}
Fig.~\ref{fig:handovernumber} shows the number of different algorithms' triggered handover times against the number of simulations increases. 
In the proposed algorithm, $\sigma$ is set as 12.9.
From Fig.~\ref{fig:handovernumber}, we can see that 
the proposed FAHP-based algorithm with location information yields reductions up to 39.18\%, 29.12\% and 10.85\% in the handover times compared with RSS-based, MCGDM and FTOPSIS algorithms, respectively. 
Compared with the FAHP algorithm without location information, the proposed algorithm reduces the handover times by 20.61\%.
In Fig.~\ref{fig:handovernumber}, we can also see that as $\sigma$ increases, the number of handover times decreases because frequent moving situations in small areas are recognized as ping-pong moving pattern and are removed from the handover trigger conditions.
    As $\sigma$ continues increasing to more than 50, the handover times remain constant. This is because, in our simulation,  nearly 81\% of network handovers are caused by network variation rather than mobility.
Therefore, with the thresholds $\sigma$ and $\delta$, the unnecessary handovers are removed efficaciously, thus avoiding ping-pong effect and  wastes of network resources. 
In Fig.~\ref{fig:handovernumber}, we can also see that FAHP is robust to location errors in terms of handover times.
\color{black}

\begin{figure}[thbp!]
    \centering
    \vspace{-0.3cm}
    \subfigure[Throughputs of MPTCP subflows]{
        \label{fig:throughputs of subflows}
        \includegraphics[width=3.6 in,height=1.2 in]{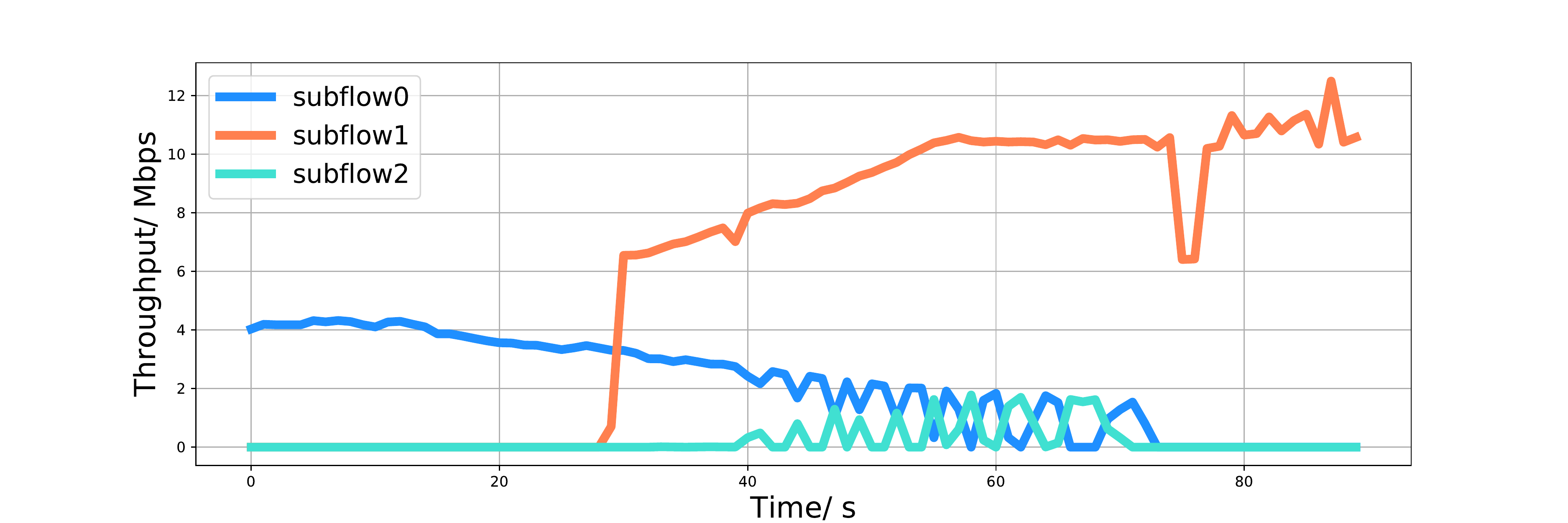}}
    \subfigure[Throughputs of MPTCP flows and TCP flows]{    \label{fig:dataflows}
        \includegraphics[width=3.6 in,height=1.2 in]{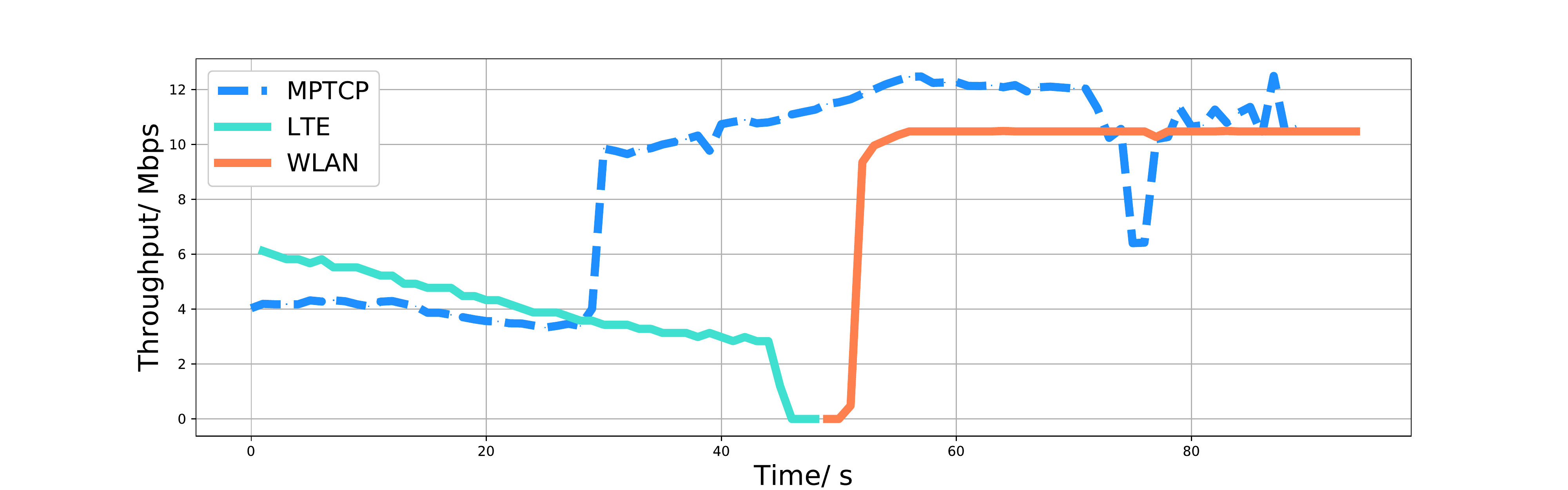}}
    \caption{Throughputs of data flows as the MMT moves.
     The MMT moves from~(1570,0) to~(1430,0) at a speed of~2~\rm m/s.}
    \label{fig:seamless}
    \vspace{-0.4cm}
\end{figure}

Fig.~\ref{fig:seamless} shows how the dataflow throughputs change as the MMT executes network handover from LTE to WLAN in SDN. 
In this experiment, we simulate different types of HetNets by modifying the parameters of network access points in Mininet-WiFi.
In Mininet-WiFi, we set up a network scenario with one LTE and one WLAN, as shown in Fig.~\ref{fig:moving_scenario}, where the LTE base station is located at~(1850,0) with a coverage radius of 400~\rm{m}, and the access point of WLAN is located at~(1400,0) with a coverage radius of 150~\rm m. 
 The MMT moves from the LTE area to the WLAN area, during which a handover is executed in the overlapped area instructed by SDN controller.

In Fig.~\ref{fig:throughputs of subflows}, we show how the MPTCP subflow throughputs vary as the MMT moves. 
From Fig.~\ref{fig:throughputs of subflows}, we can see that  
as the distance between the MMT and LTE base station increases, the throughputs of subflow0 in MPTCP0 decrease for the decaying RSS from LTE. 
When the MMT moves into the overlapped area of LTE and WLAN, SDN controller perceives the handover condition and sends the MMT handover signaling to initialize MPTCP1, whose subflow1 and subflow2 will connect to WLAN and LTE, respectively.
In Fig.~\ref{fig:throughputs of subflows}, we can also see that the decrease of subflow0 and subflow2 fluctuates in a complementary manner, this is because they share the fading bandwidth from LTE.
As the MMT moves into the area that only WLAN covers, subflow0 and subflow2, which both connect to LTE, are terminated, while subflow1, which connects to WLAN, maintains its throughputs. 
In this way, the service continuity during handover in this SDHetNet is guaranteed.

In Fig.~\ref{fig:dataflows}, we compare the proposed MPTCP-based handover execution with the conventional handover method, which uses one interface in the same scenario. 
In Fig.~\ref{fig:dataflows}, the turquoise line and the orange line represent the throughputs of dataflows using one interface through LTE and WLAN, respectively. 
In Fig.~\ref{fig:dataflows}, we can see that the conventional handover using one interface has an inevitable interruption of connection during handover. 
While with two interfaces simultaneously transmitting data through the HetNets, the MPTCP-based handover mechanism maintains its connection and increases data throughputs during the handover process. 
Clearly, the proposed method realizes a seamless handover and improves the quality of user experience.

\section{Conclusion}
In this paper, we have proposed a \color{black} mobility-aware \color{black} seamless handover method with mobility information and MPTCP in SDHetNets. 
First, for network discovery
we developed an ESN model to determine the user's next access candidate networks by predicting the mobile user's location.
Next, for network selection, we proposed an FAHP-based network selection algorithm, for which the target network is selected with the highest QoE metric considering user preferences, QoS requirements and real-time network attributes.
The proposed algorithm also recognizes the ping-pong moving pattern of users and removes the mobility pattern from the handover trigger conditions.
Then, for handover execution, we proposed a signaling process that uses MPTCP to transmit data through multiple HetNets during handover in SDN.
The simulation results have demonstrated that the proposed method realizes a seamless handover and yields a significant improvement in terms of handover times and service continuity compared to conventional methods.

\ifCLASSOPTIONcaptionsoff
  \newpage
\fi

\def\baselinestretch{0.95}



\bibliographystyle{IEEEtran}

\vspace{12pt}

\end{document}